\documentclass[preprint,12pt]{elsarticle}
\usepackage{amsmath,amsfonts,amsthm,commands,graphicx,amssymb}
\def\vK{von K\'{a}rm\'{a}n }
\def\i{\textrm{i}}

\begin{document}
\begin{frontmatter}
\title{On the noise generation and unsteady performance of combined heaving and pitching foils}

\author[psu]{Nathan Wagenhoffer}
\ead{wagenhoffer@psu.edu}
% \affiliation{ Mechanical Engineering and Mechanics, Lehigh University}
\author[leh]{Keith W Moored}
% \affiliation{ Mechanical Engineering and Mechanics, Lehigh University}
\author[leh]{Justin W Jaworski}
\address[psu]{Engineering, Pennsylvania State University Abington}
\address[leh]{Mechanical Engineering and Mechanics, Lehigh University}

%**********
% Abstract
%**********

\begin{abstract}
  A transient two-dimensional acoustic boundary element solver is coupled to a potential flow boundary element solver via Powell's acoustic analogy to determine the acoustic emission of isolated hydrofoils performing biologically-inspired motions. The flow-acoustic boundary element framework is validated against experimental and asymptotic solutions for the noise produced by canonical vortex-body interactions.  The numerical framework then characterizes the noise production of an oscillating foil, which is a simple representation of a fish  caudal fin. A rigid NACA 0012 hydrofoil is subjected to combined heaving and pitching motions for Strouhal numbers ($0.03 < St < 1$) based on peak-to-peak amplitudes and chord-based reduced frequencies ($0.125 < f^* < 1$) that span the parameter space of many swimming fish species. A dipolar acoustic directivity is found for all motions, frequencies, and amplitudes considered, and the peak noise level increases with both the reduced frequency and the Strouhal number. A combined heaving and pitching motion produces less noise than either a purely pitching or purely heaving foil at a fixed reduced frequency and amplitude of motion.  Correlations of the lift and power coefficients with the peak root-mean-square acoustic pressure levels are determined, which could be utilized to develop long-range, quiet swimmers.
\end{abstract}

\end{frontmatter}

%*********
% Chapter
%*********
\section{Introduction}
Many aquatic animals oscillate their fins to propel themselves quickly through the ocean.  Their propulsion is based on an unsteady flow paradigm that is distinct from the steady flow paradigm that underpins the design of most man-made underwater vehicles.  Consequently, animals can excel at maneuvering and rapid accelerations \cite{lauder2015fish,sfakiotakis1999review} in addition to high-speed and high-efficiency swimming \cite{Tri1,Tri2}.  This multi-modal performance has spurred vigorous research over the last two decades into bio-inspired autonomous underwater vehicles (BAUVs) \cite{triantafyllou1995efficient,liu20043d,hu2006biologically,clapham2015developing}, which would possess the additional benefit of exceptional \textit{stealth} \cite{bandyopadhyay2005trends,moored2011batoid}.  It is generally expected that fish-like swimming motions lead to quiet acoustic signatures \cite{bandyopadhyay2005trends}, which (even if detected) would resemble the noise signature of real fish, making a BAUV exceptionally difficult to detect and identify \cite{wagenhoffer2017accelerated}.  However, to date the noise signatures of swimming animals and BAUVs have not been adequately quantified, nor have the trade-offs between the noise signature and performance of unsteady swimmers been examined. The competition between acoustic stealth and fluid dynamic performance also emerges for biologically-inspired aerial vehicles \cite{jaworski2013aerodynamic,clark2017bioinspired,hajian2017steady,jaworski2020aeroacoustics} that seek novel means to suppress flow-noise generation.  

Most sounds produced by fish result from aggressive actions, spawning, or reproductive behavior \cite{fay2009fish}. The noise made during these actions can be categorized into two broad  categories: active acoustic signaling through morphological structures and passive noise generated during swimming, feeding, or respiration. The actively-produced noises include vocal calls, swimbladder motions, and drumming \cite{ladich2006sound,luczkovich2002using}. Active fish signaling is not of interest here, as there exist quantified data for several species \cite{ladich2006sound}. However, fish produce a vortex wake by simply oscillating their fins during swimming, which is a well-known source of noise \cite{fay2009fish}. Yet, the locomotive noise of fish during steady-state rectilinear swimming remains relatively unquantified. Fish locomotion produces low-amplitude hydrodynamic noise that is challenging to record reliably and has received limited attention in the literature~\cite{fay2009fish}. 

Instead of examining live fish, numerical tools can be used to simulate the motions of fish and their resulting unsteady vortical wakes to estimate their noise production.  For instance, one approach to simulating the unsteady flow field around a swimming fish is the boundary element method (BEM) \cite{moored2018unsteady}.  This potential flow method discretizes surfaces representing shear layers in the flow, i.e. a solid body surface and the wake, into a collection of elements \cite{hess1967calculation,katz2001low}. This method approximates the flow as incompressible, inviscid, and irrotational (except on the elements).  Additionally, the reduction  of the solution domain to solve only for the strengths of elements on the boundaries enables the rapid simulation of unsteady flow phenomena.  Acoustic BEMs have  been used to determine the far-field noise of an airfoil in a turbulent flow, such as by Glegg and Devenport~\cite{glegg2010panel}.  Their method predicts the frequency-domain acoustic far field based on a form of the surface loading due to the energy spectrum of the turbulent boundary layer immediately upstream of the trailing edge. However, this method is restricted to single airfoils in steady flows and does not account for interactions that would be present in many-bodied systems of swimmers or fliers.

Once an unsteady flow field is simulated, an acoustic analogy is a common and effective tool to post-process the flow data and determine the noise production \cite{wang2006computational,karimi2016trailing}. Lighthill~\cite{lighthill1952sound} first developed an acoustic analogy by recasting the Navier-Stokes equations into a wave equation in terms of perturbations of the fluid density. A nearly ubiquitous approach to predicting flow noise,  Lighthill's acoustic analogy has become central to many aeroacoustic studies \cite{wang2006computational,karimi2016trailing,wang1996computation,wolf2011trailing}. For example, Powell~\cite{powell1964theory} adapted the Lighthill analogy to consider  all of the unsteady flow fluctuations as vorticity and designated this vorticity as the forcing function for the acoustic system.
% The use of an unsteady flow model coupled to an acoustic analogy can give an approximation to the type of sound production associated with locomotion. The coupling will allow an investigation to define the difference between biological systems and their mechanical counterparts. 
Vorticity also plays a central role in the  noise theory by Howe~\cite{howe_1975} that is adapted from Lighthill's formulation.

Motivated by these observations, the present study makes two principal technical advancements.  First, a coupled flow-acoustic BEM is  developed to predict the noise generation and performance of hydrofoils in unsteady motion that can also account for multi-body interactions, such as those that occur in fish schools.  Second, the novel flow-acoustic BEM will be used to examine the performance and noise production of a hydrofoil in combined heaving and pitching motions, which acts as a simple, two-dimensional representation of an oscillating fish caudal fin.  This paper is laid out across three main sections. The first section describes the potential flow and acoustic boundary element methods used in the current study. Next, the coupling between the potential flow and acoustic solvers is presented along with validations of the solver against available  experimental data and asymptotic solutions of vortex-body interaction noise. Finally, the performance and noise production of a combined heaving and pitching hydrofoil are examined with respect to non-dimensional frequency, amplitude, and heave-to-pitch ratio. 

\section{Potential Flow Boundary Element Method}\label{S:flow_bem}

\begin{figure}
    \centering
    \includegraphics[width=0.9\textwidth]{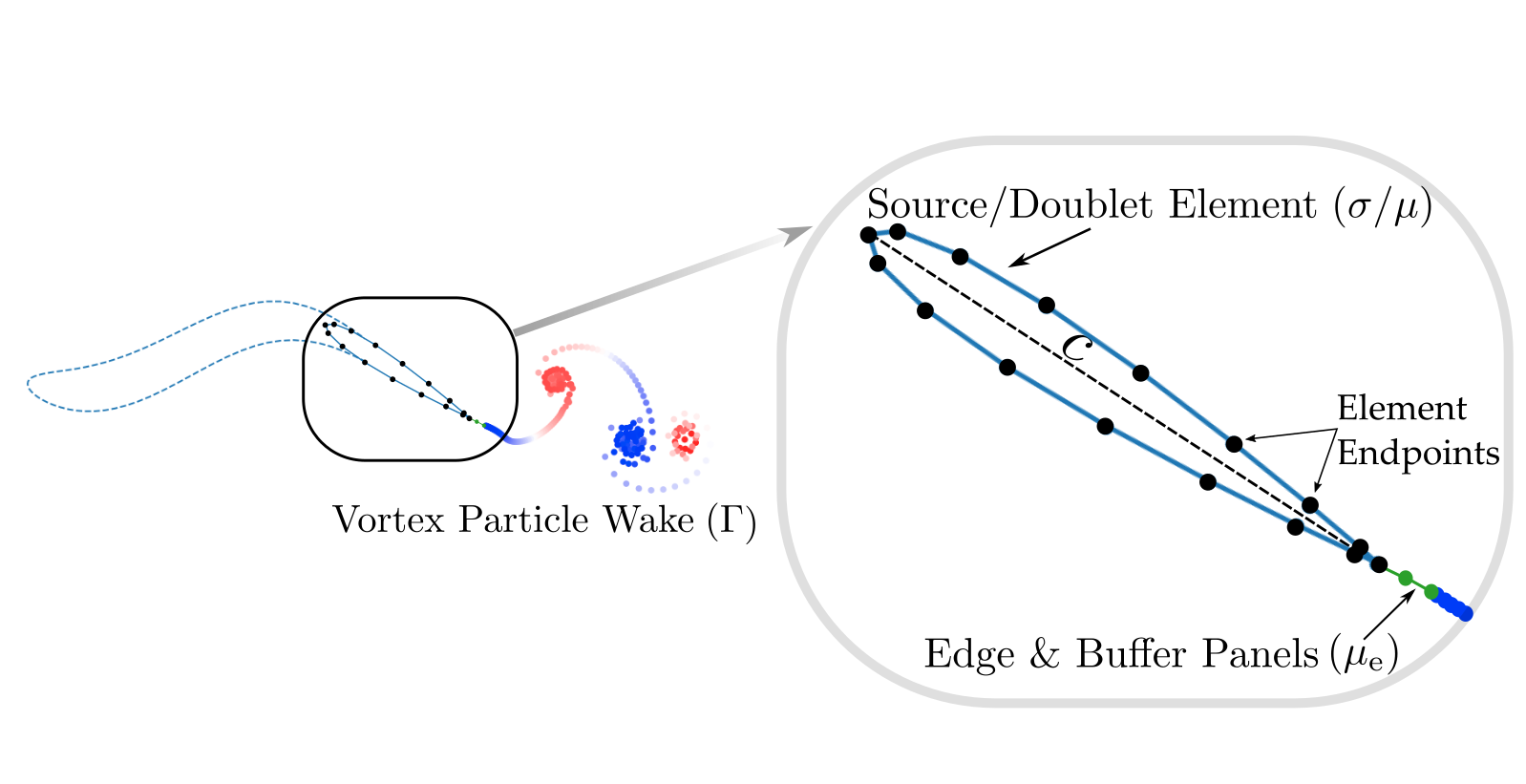}
    \caption{Schematic of a fish with a blown up section depicting the propulsor as the discrete geometry for the BEM. A propulsor of chord length $c$ is discretized with source/doublet elements (blue lines with endpoints indicated by black circles). The edge and buffer panels (green lines with endpoints indicated by green circles)  are connected to the trailing edge of the propulsor. The vortex particle wake (blue/red circles) is shed from the edge/buffer panels. }
    \label{fig:heaving_discrete_propulsor}
\end{figure}

This section details the two-dimensional unsteady boundary element method used in this work. The potential flow solver  is an adaptation of the panel method described by \citeauthor{willis2007combined} \cite{willis2007combined}. The inviscid flow around hydrofoils can be found by solving the Laplace equation with an imposed no-penetration  boundary condition on the surface,
\begin{equation}
    \nabla \phi \cdot \mathbf{\hat{n}} = 0 ~~ {\rm on}~~ S_{\rm{b}},
\end{equation}
where $ \mathbf{\hat{n}}
$ is the outward unit normal of the surface. The boundary integral equation integrates the effects of the combined distribution of sources and doublets on the body surface $S_{\rm{b}}$ and doublets on edge panel $S_{\rm{e}}$, with vortex particles in the wake $S_{\rm{w}}$ (cf.\ Fig.~\ref{fig:heaving_discrete_propulsor}). The scalar potential may be written as % A Helmholtz decomposition separates the flow field into scalar $\phi$ and vector potentials $\mathbf{\Psi}$. 
\begin{equation}\label{eqn:flowBIE}
\phi(\mathbf{t}) = \int_{S_{\rm{b}}} \left[\sigma(\mathbf{s}) G(\mathbf{t},\mathbf{s}) - \mu(\mathbf{s})\hat{\mathbf{n}}\cdot \nabla G(\mathbf{t},\mathbf{s})\right]\rm{d} S -\int_{S_{\rm{e}}}\mu_{\rm e} (\mathbf{s})\hat{\mathbf{n}}\cdot \nabla G(\mathbf{t},\mathbf{s})\rm{d} S,
\end{equation}
where  $\mathbf{s}$ is a source location, $\mathbf{t}$ is the observer location, and $G(\mathbf{t},\mathbf{s})= \frac{1}{2\pi}\ln|\mathbf{t}-\mathbf{s}|$ is the two-dimensional Green's function for the Laplace equation. The source and doublet strengths are defined respectively as
\begin{eqnarray}
\sigma = \hat{\mathbf{n}} \cdot ( \mathbf{U} + \mathbf{U}_{\rm{rel}} - \mathbf{U}_\omega),\\
\mu = \phi_I - \phi, 
\end{eqnarray}
where $\mathbf{U}_\omega$ is velocity induced by the vortex particles in the field, $\mathbf{U}$ is the body velocity, $\mathbf{U}_{\rm rel}$ is the velocity of the center of each element relative to the body-frame of reference, and $\phi_I$ is the interior potential of the body.  
At each time-step, vorticity is defined at the trailing-edge to satisfy the Kutta condition. The trailing edge panel is assigned the  potential difference between the upper and lower panels at the trailing edge of the foil, $\mu_{\rm e} = \mu_{\rm{upper}}-\mu_{\rm{lower}}$. Application of the Kutta condition ensures that the vorticity at the trailing edge of the hydrofoil is zero. An implicit Kutta condition for the two-dimensional flow solver is similar to the methods mentioned in \cite{willis2007combined,jones1997numerical}. The iterative implicit Kutta condition employs a Newton's method to define the length and angle of the trailing edge panel in order to minimize the pressure across the trailing edge. The evolution of vorticity in the domain is governed by
\begin{equation}
\frac{\partial \boldsymbol{\omega}}{\partial t} + \mathbf{U}\cdot\nabla\boldsymbol{\omega}= \boldsymbol{\omega}\cdot\nabla \mathbf{U},
\end{equation}
where the vorticity field $\boldsymbol{\omega}$ is represented by discrete, radially-symmetric, desingularized Gaussian vortex particles. The induced velocity of the vortex blobs is determined by the Biot-Savart law \cite{cottet2000vortex},

\begin{equation}\label{eqn:biot-sav}
\mathbf{u}(\mathbf{t},t) = \sum_{i=1}^N \frac{\Gamma_i}{2\pi} \left(1-\exp\left(\frac{-|\mathbf{t - s}|}{2{r_{\rm{cut}}}^2}\right)\right),\\
\end{equation}
where $\Gamma_i$ is the circulation of the $i$th vortex particle, and $r_{\rm{cut}}$ is the cut-off radius. Following the work of Pan \textit{et al}. \cite{pan2012boundary}, the cut-off radius is set to $r_{\rm{cut}} = 1.3\Delta t$ for time step $\Delta t$ to ensure that the wake particle cores overlap and a thin vortex sheet is shed. The evolution of the vortex particle position is updated using a forward Euler scheme \cite{willis2007combined},
\begin{equation}
\mathbf{x}(t+1) = \mathbf{x}(t) +\mathbf{u}(\mathbf{x}(t),t)\Delta t.
\end{equation}

The use of discrete vortices to represent the wake instead of panels alters the process of shedding vorticity into the wake in comparison to more classical source-doublet methods \cite{katz2001low}. Two edge panels are set behind a foil. The first edge panel, set with the empirical length of $l_{\rm{panel}} = 0.4 U_{\infty}\Delta t$ \cite{katz2001low}, satisfies the Kutta condition at the trailing edge. Next, the buffer panel is attached to the edge panel and stores information about the previous time step. Figure \ref{fig:heaving_discrete_propulsor} illustrates the distinction of the source/doublet panels across a foil, the arrangement of the trailing-edge and buffer panels, and how the vortex particle wake behaves behind the body.

The vortex particles influence on the body is accounted for in the definition of the source strength $\sigma$. The vortex particle induced velocity also augments the pressure calculation put forth by \citeauthor{katz2001low} \cite{katz2001low}. The surface pressure is determined by 
\begin{equation}\label{eqn:flow-pressure}
    \frac{P_\infty - P(x)}{\rho} = \left.\frac{\partial \phi_{\rm{wake}}}{\partial t}\right|_{\rm{body}} +\left.\frac{\partial \phi}{\partial t}\right|_{\rm{body}} - (\mathbf{U}+\mathbf{u}_{\rm_{rel}})\cdot(\nabla \phi + \mathbf{U}_\omega)) +\frac{1}{2}|\nabla \phi + \mathbf{U}_\omega)|^2,
\end{equation}
where $\partial \phi_{\rm{wake}}/\partial t = \Gamma \dot{\theta}/(2\pi)$ is the time rate of change due to a vortex particle with circulation $\Gamma$ at an angle $\theta$ from the observation point. Equation (\ref{eqn:flow-pressure}) is similar to the form put forth by \citeauthor{willis2007combined} \cite{willis2007combined}, but here  $\partial \phi_{\rm{wake}}/\partial t$ is the positional change of a vortex with respect to a panel and does not require the solution of a secondary system to find the influence of wake vortices onto the body surface. The appendix presents validation of this methodology against available theoretical and numerical results.

\section{Acoustic Boundary Element Method}\label{c:acoustic_bem}

The Helmholtz wave equation for a homogeneous medium and its boundary conditions are 
\begin{eqnarray}
\nabla^2\phi+\kappa^2\phi = 0\label{eqn:ap2helm},   \\
\phi(\mathbf{x}) = d_n  ~~ {\rm on}~~ S_{\rm{b}}, \\
\frac{\partial \phi}{\partial n}(\mathbf{x}) = g_n  ~~ {\rm on}~~ S_{\rm{b}},
\end{eqnarray}
where $\phi$ is the time-independent velocity potential, $\kappa = 2\pi f/c_0$ is the acoustic wavenumber, $c_0$ is the speed of sound, $d_n$ is a Dirichlet boundary condition, and $g_n$ is a Neumann boundary condition. 

Application of Green's second identity to Eq.~(\ref{eqn:ap2helm}) moves all of the information of the system onto the boundary:
\begin{equation}\label{eqn:ac1BIE}
    a(\mathbf{t})\phi(\mathbf{t}) = \int_{S_{\rm{b}}}\left[\frac{\partial G(\mathbf{t},\mathbf{s})}{\partial n}\phi(\mathbf{s}) -G(\mathbf{t},\mathbf{s})\frac{\partial \phi}{\partial n}(\mathbf{s})\right]\rm{d} S_{\rm{b}},
\end{equation}
where $a(\mathbf{t})= \frac{1}{2}$ on the boundary and $a(\mathbf{t})= 1$ in the exterior field.  The two-dimensional acoustic Green's functions are
\begin{eqnarray}
    G(\kappa,\mathbf{t},\mathbf{s}) = \frac{\i H_0^{(1)}(\kappa|\mathbf{t}-\mathbf{s}|)}{4}, \\
    \frac{\partial G}{\partial n}(\kappa,\mathbf{t},\mathbf{s}) = \frac{-\i \kappa H_1^{(1)}(\kappa |\mathbf{t}-\mathbf{s}|)}{4},
\end{eqnarray}
which correspond respectively to an acoustic monopole and dipole, and $H_n^{(1)}$ is the Hankel function of the first kind of order $n$. In the remainder of this work, $G(\kappa,\mathbf{t},\mathbf{s}) = G(\mathbf{t},\mathbf{s})$, as the wavenumber is constant for each solution. 

Differentiation of Eq.~(\ref{eqn:ac1BIE}) with respect to the outward normal of the boundary produces a quadrupole system,
\begin{equation}\label{eqn:acquad}
    \frac{\partial\phi(\mathbf{t})}{\partial n} = \int_{S_{\rm{b}}}\frac{\partial^2 G(\mathbf{t},\mathbf{s})}{\partial n_{\mathbf{t}} \partial n_{\mathbf{s}}}\phi(\mathbf{s})\rm{d} S_{\rm{b}},
\end{equation}
where $n_{\mathbf{s}}$ and $n_{\mathbf{t}}$ are the outward normals at the source and observer, respectively. 

A combination of Eqs.~(\ref{eqn:ac1BIE}) and (\ref{eqn:acquad}) for points on the boundary arrives at the Burton-Miller formulation \cite{wolf2011trailing,kirkup2007boundary}:
\begin{equation}\label{eqn:BurtonMiller}
    \int_{S_{\rm{b}}}\left(\frac{\partial G(\mathbf{t},\mathbf{s})}{\partial n}\phi(\mathbf{s}) +\frac{1}{2}\phi(\mathbf{s})\right)\rm{d} S_{\rm{b}} + \beta\int_{S_{\rm{b}}}\frac{\partial^2 G(\mathbf{t},\mathbf{s})}{\partial n_{\mathbf{t}} \partial n_{\mathbf{s}}}\phi(\mathbf{s})\rm{d} S_{\rm{b}} = \phi(x) +\beta \frac{\partial\phi(\mathbf{t})}{\partial n},
\end{equation}
where $\beta = \i/\kappa$ is a chosen coupling parameter \cite{wolf2011trailing}. The frequency domain problem (\ref{eqn:BurtonMiller}) is converted into a transient solution by the application of the convolution quadrature method, which is detailed in the next section. 

\subsection{Time discretization}
The frequency potential operators in Eq.~(\ref{eqn:BurtonMiller}) are evaluated as convolution integrals. Green's functions in the frequency domain problem are Laplace transforms  of the retarded-time Green's function, which permit their  convolution with the potential field. The potential field is evaluated by a convolution quadrature. 
This methodology of time discretization can be achieved via a convolution quadrature method put forth by Lubich \cite{Lub}. A representative example of a convolution system is
\begin{eqnarray*}
    \int_0^t f(t-\tau)\phi(\tau)d\tau = g(t).
\end{eqnarray*}
Here $f$ represents a retarded-time operator, a characteristic differential operator of the transient wave equation, $\phi$ is  some known potential distribution, and $g(t)$ is a transient forcing function. The interested reader may consult Hassell and Sayas \cite{hassell2016convolution} for a detailed explanation of the convolution quadrature method. The retarded-time operator is a convolution that can be discretized by splitting the time domain into $N+1$ time steps of equal spacing, yielding $\Delta t = T / N$ and $t_n = n\Delta t$ for $n = [0,1,...,N]$. The discrete convolution can be viewed as a sum of weights of the $F$ operator at discrete times of $\phi$:
\begin{eqnarray}
	F  \frac{\partial \Phi(t_n)}{\partial t}^{\Delta t}  = \sum_{j=0}^n w_{n-j}^{\Delta t}(F)\phi^{\Delta t}(t_j),
\end{eqnarray}
where $F$ represents the Laplace transform of the $f$ operator, and the superscript $\Delta t$ indicates the weight for a specific time-step size. 
The series expansion can be arranged to solve for the convolution weights, $w$:
\begin{eqnarray}
	F\left(\frac{\gamma(\zeta)}{\Delta t}\right) = \sum_{n=0}^\infty w_{n-j}^{\Delta t} \zeta^n, \quad |\zeta|<1, \\
	w_{n-j}^{\Delta t} = \frac{1}{2\pi i}\oint_C \frac{F(\frac{\gamma(\zeta)}{\Delta t})}{\zeta^{j+1}}d\zeta,
\end{eqnarray}
where $C$ is a circle of radius $0 < \lambda < 1$ centered at the origin. A second-order backwards difference function, $\gamma(\zeta) = (1- \zeta) + \frac{1}{2}(1-\zeta)^2$, is used to define the spacing of the integration. A review of other integration methods that can be incorporated into the convolution quadrature method is presented in  \citeauthor{hassell2016convolution} \cite{hassell2016convolution}. Employing a scaled inverse transform, the weights become
\begin{eqnarray}\label{eqn:weighted}
	w_{n-j}^{\Delta t,\lambda}(F) = \frac{\lambda^{-j}}{N+1}\sum_{l=0}^N F(s_l) \zeta_{N+1}^{lj},
\end{eqnarray}
where $\zeta_{N+1} = \exp\left(\frac{2\pi i}{N+1}\right)$ is the discrete Fourier transform scale, and $s_l = \gamma(\lambda\zeta_{N+1}^{-l}) /\Delta t$ is the accompanying time-dependent complex wavenumber.  The value of $s_l$ is different for each time step and provides the link between the frequency-domain solver and a transient boundary integral equation. For this formulation, $\lambda = \Delta t ^{3/N}$ is selected based on the error analysis of Banjai and Sauter \cite{Banjai08}.

Placing (\ref{eqn:weighted}) into the  boundary value problem (\ref{eqn:BurtonMiller}) yields a system of $N+1$ equations,
\begin{eqnarray}\label{freqProbs}
	\frac{\lambda^{-j}}{N+1}\sum_{l=0}^N F(s_l,\mathbf{x})\hat{\phi_l}(\mathbf{x}) \zeta_{N+1}^{lj}    = \frac{\lambda^{-j}}{N+1} \sum_{l=0}^N \hat{g_l}\zeta^{lj}_{N=1},
\end{eqnarray}
where $F$ is the linear combination of operators on the left hand side of Eq.~(\ref{eqn:BurtonMiller}). Here $g_n$ is a discrete representation of the mixed boundary condition. The inverse discrete Fourier transform of the convolution is
\begin{eqnarray}\label{eqn:DFF}
   	\hat{\phi_l} =  \sum_{j=0}^N \lambda^j\phi_j^\lambda\xi_{N+1}^{-lj},
\end{eqnarray}
which produces the transient solution. 

In summary, the convolution quadrature method \cite{Banjai08} discretizes a transient wave problem  into a system of frequency-domain (Helmholtz) wave equations that are uncoupled in time. This discretization allows  $N+1$ independent solutions of  Eq.~(\ref{eqn:BurtonMiller})  in the frequency domain using wavenumbers $s_l$ that are generated via the convolution quadrature method. The time-domain solution is recovered by applying the inverse Fourier transform (\ref{eqn:DFF}).

\section{Acoustic Analogies}
%%History and definition of analogies
The Lighthill acoustic analogy~\cite{lighthill1952sound} is an exact rearrangement of the Navier-Stokes equation, where the resulting wave equation is forced by the so-called Lighthill tensor, $T_{ij}$. For example, $T_{ij}$ may be used as a forcing function in Eq.~(\ref{eqn:ap2helm}). The flow is assumed inviscid, without thermal losses, and at low Mach number $(M^2 \ll 1)$. The form of the Lighthill tensor under these conditions is reduced to the Reynolds stress contribution, $T_{ij} \approx \rho u_i u_j$ \cite{howe2003theory}. Taking two spatial derivatives of this tensor yields the quadrupole source for Lighthill's acoustic analogy. This quadrupole is related to the vorticity field by
\begin{equation}
\frac{\partial^2u_i u_j}{\partial x_i \partial x_j} = \nabla \cdot (\boldsymbol{\omega} \times \mathbf{u}) + \nabla^2\left(\frac{1}{2}u^2\right),
\end{equation}
where $u^2 = \mathbf{u} \cdot \mathbf{u}$. The Powell acoustic analogy, a derivative of the Lighthill analogy, uses this form of the velocity field as its forcing function \cite{powell1964theory}.

%Powells Definitions 
The Powell acoustic analogy allows the direct relation of the vorticity defined by the flow solver to be the forcing function of the acoustic solver. The present flow solver defines the vorticity that is bound to the body and shed into the wake. Other vortical noise sources, such as the broadband content in a turbulent boundary layer, may also be incorporated but are neglected in the present work to focus on the acoustic interactions involving incident and shed vorticity. The Powell acoustic analogy states that in free space the forcing of the wave equation is a function of the vorticity in the field,
\begin{eqnarray}
    \partial_t^2 P-\nabla^2 P = \nabla \cdot(\boldsymbol{\omega} \times \mathbf{u}),\label{eqn:powell}\\
    \frac{\partial G}{\partial n} = 0 \quad {\rm on} \quad S_{\rm{b}}.\label{eqn:powell_noflux}
\end{eqnarray}
Using a Green's function solution, and applying an integration by parts, the pressure in the field is determined by
\begin{equation}\label{eqn:vort_pres}
    P(\mathbf{t},t) = \rho \int_{S_{\rm{b}}}(\boldsymbol{\omega} \times \mathbf{v})\cdot \frac{\partial G}{\partial n_{\mathbf{s}}}\rm{d} S_{\rm{b}},
\end{equation}
where $\partial G /\partial n_{\mathbf{s}}$ is the two-dimensional potential flow Green's function. The pressure integral in Eq.~(\ref{eqn:vort_pres}) is applicable regardless of whether or not a solid body is present \cite{kambe1986acoustic}. The use of the potential flow Green's function  imposes instantaneously the vortical acoustic loading onto a solid body, as in the asymptotic methods of \citeauthor{kambe1986acoustic} and \citeauthor{kao2002body} \cite{kambe1986acoustic,kao2002body}. Also, the use of the flow Green's function instead of the retarded potential acoustic Green's function \cite{kambe1986acoustic}, i.e., $G = \frac{H(t-|\mathbf{t}|/c_0)}{2\pi\sqrt{t-|\mathbf{t}|/c_0}}$, removes any singularities in time in addition to the slow decay tail created by the Heaviside function found in the numerator. For all of the flow scenarios observed in this work, the small products of the acoustic wavenumber with the chord length (Helmholtz number) and with the distance between dominant wake vortices with the foil in low-Mach-number locomotion ensures acoustically compact vortex-body interactions~\cite{howe2003theory}.
%BOundary condition definitions
Equation (\ref{eqn:vort_pres}) provides the pressure at an arbitrary point in space, for instance the collocation point of a discrete foil. This approach is sufficient to solve a Dirichlet boundary condition, but Eq.~(\ref{eqn:powell_noflux}) is a no-flux Neumann boundary condition. The Neumann boundary condition can be satisfied by arranging Euler's equation to define the acoustic dipole potential needed to guarantee no fluid flux through the surface of the discrete geometry, 
\begin{equation}\label{eqn:Euler}
    \rho \frac{\partial \mathbf{u}}{\partial t} = -\nabla P.
\end{equation}
In Eq.~(\ref{eqn:Euler}) the velocity is taken to be the normal induced velocity from each discrete vortex in the domain, including the discrete vortex particles that comprise the wake and vortex values distributed along the foil found via Eq.~(\ref{eqn:flowBIE}). A rearrangement of Eq.~(\ref{eqn:Euler}) that considers the outward normal pressure on the foil results in
\begin{equation}\label{eqn:Eulerdpdn}
    \frac{\partial P}{\partial n}(\mathbf{t}) = -\rho \sum_{i=0}^N   \frac{\partial( \mathbf{u}_i\cdot \hat{\mathbf{n}}(\mathbf{t}))}{\partial t}.
\end{equation}
The pressure in Eq.~(\ref{eqn:vort_pres}) and its normal derivative in Eq.~(\ref{eqn:Eulerdpdn}) are the boundary conditions to the Burton-Miller acoustic BEM formulation in Eq.~(\ref{eqn:BurtonMiller}).

\subsection{Validation of Acoustic Analogy for Vortex-Body Interactions}\label{S:self-noise-valid}
%Validation prob description
The Powell acoustic analogy is now validated as a suitable forcing function for the one-way coupled flow-acoustic BEM. First, the  canonical problem by Crighton \cite{crighton1972radiation} whereby a line vortex generates noise as it passes round a half-plane is used to verify the flow-acoustic BEM for edge scattering. Given a point vortex that advects around a half-plane lying at $y=0$ with the edge located at the origin, the vortex path is described analytically by
\begin{equation*}
     r = \ell \sqrt{1 +\left(\frac{Ut}{\ell}\right)^2}, \quad~\quad \theta = 2\tan^{-1} \left(-\frac{U t}{\ell} \right),
\end{equation*}
where $r$ and $\theta$ are the radial components of the vortex position. The characteristic vortex velocity is $U = \Gamma/8\pi \ell$, with $\ell$ being the distance of closest approach of the vortex to the trailing edge. Howe~\cite{howe2003theory} (see also \cite{howe_1975}) determined the time-varying magnitude of the acoustic pressure produced by the vortex motion near the edge using an approximate Green's function for a half plane with the observer in the acoustic far field:
\begin{align}\label{eqn:howesCrighton}
    P(\mathbf{t},t) &\approx \frac{\rho \Gamma^2}{4\pi \ell^2}\left(\frac{\ell}{|\mathbf{t}|}\right)^{\frac{1}{2}}\sin\left(\frac{\theta}{2}\right)\left[   \frac{\Gamma t / 8 \pi \ell^2}{[1 +(\Gamma t/ 8 \pi \ell^2)^2]^{5/4}}\right]_{t-\frac{|\mathbf{t}|}{c_0}},   ~\mathbf{|t|}& \rightarrow{} \infty.
\end{align}
% \begin{align}\label{eqn:howesCrighton}
%     P(\mathbf{t},t) &\approx \frac{\rho \Gamma^2}{4\pi \ell^2}\left(\frac{\ell}{|\mathbf{t}|}\right)^{\frac{1}{2}}\sin\left(\frac{\theta}{2}\right)\left[    \frac{1}{2\sqrt{r}}\frac{\d r}{\d t}\cos\left(\frac{\theta}{2}\right) + \frac{\sqrt{r}}{2}\sin\left(\frac{\theta}{2}\right)\frac{\d \theta}{\d t}
%     \right]_{t-\frac{|\mathbf{t}|}{c_0}}, \nonumber \\
%     ~\mathbf{|t|}& \rightarrow{} \infty.
% \end{align}
%
The brackets indicate evaluation at the retarded time. Note that this model neglects any vortex shedding by the edge. 

\begin{figure}
    \centering
    \includegraphics[width=1\textwidth]{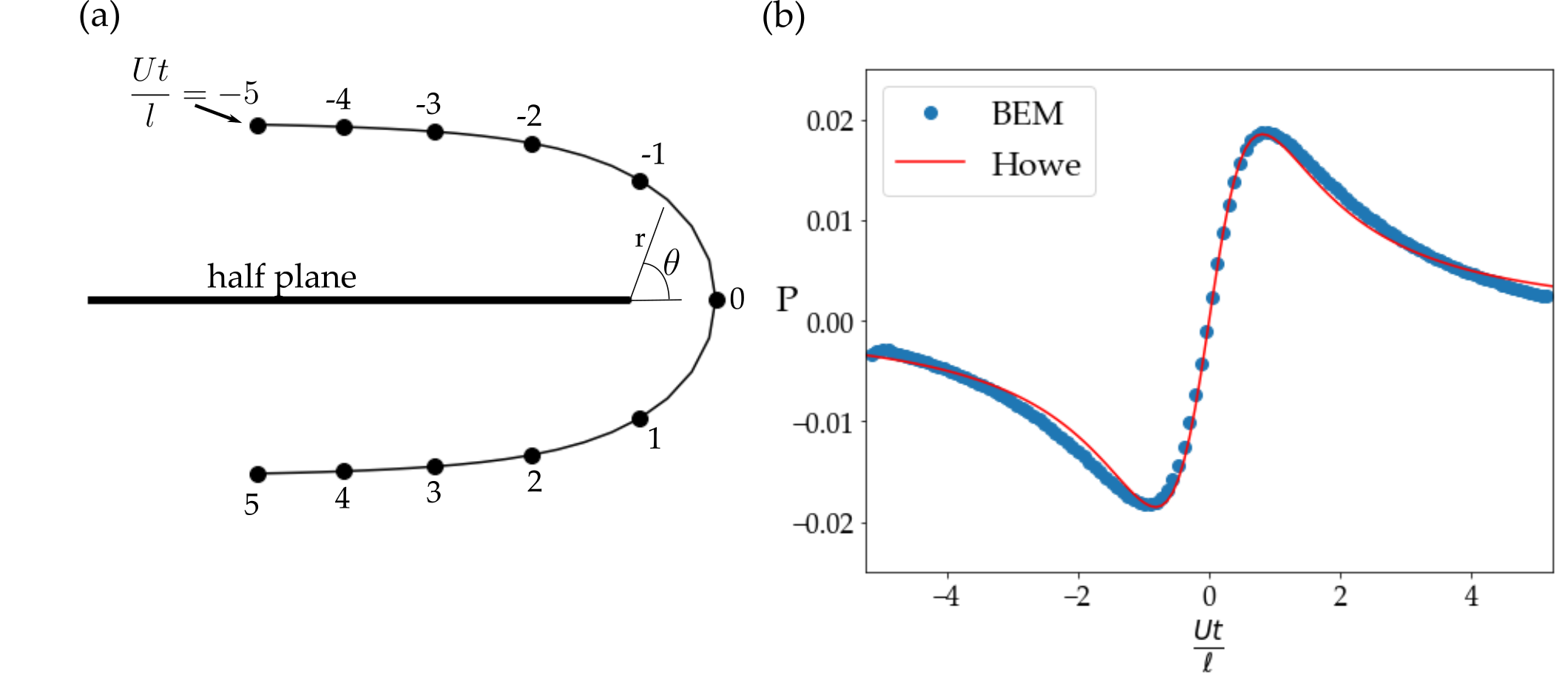}
    \caption{Sound produced by vortex motion near a half plane. (a) Vortex path round the trailing edge of a half plane. The closest position of the vortex to the body $\ell$ occurs at $Ut/\ell = 0.$ (b) Acoustic response computed by the BEM (blue circles) and from Howe's solution (red line). }
    \label{fig:crighton}
\end{figure}

Figure \ref{fig:crighton} illustrates the vortex path near the half plane and presents a comparison of the analytical acoustic response from Eq.~(\ref{eqn:howesCrighton}) against the numerical results from the BEM developed in \S\S\ref{S:flow_bem} and \ref{c:acoustic_bem}. The solution was created by approximating a half plane with a 10 m flat plate and passing a vortex of strength $\Gamma$ = 1 m$^2$ s$^{-1}$ to within a distance of $\ell$ = 0.25 m at the trailing edge in a medium with density $\rho$ = 1 kg m$^{-3}$. The observer location is $|\mathbf{t}|$ = 50 m above the trailing edge. The half plane was approximated with a 10 m flat plate discretized with 512 boundary elements in a cosine distribution which ensures a denser concentration of elements near the edge. The doubling of elements on the body from 128 to 256 elements produced less than 0.1\% change in the acoustic solution as observed at $|\mathbf{t}|$. Agreement of the acoustic responses are seen when the vortex is near the trailing edge, i.e.~$|Ut/\ell|$ < 1, and the system begins to diverge slightly outside of that range. The divergence of the solution outside of this time is likely due to the limitation in the Howe solution that the loading is only at the trailing edge, where as the BEM implicitly solves for loading along the flat plate as the vortex continues to travel near the body. The agreement between the BEM and Howe's solution verifies the accuracy of the acoustic portion of the BEM in modeling trailing edge scattering. The fully coupled solver including the unsteady potential flow portion can be examined further for vortex-body interaction cases.

\begin{figure}[!ht]
    \centering
    \includegraphics[width=0.9\textwidth]{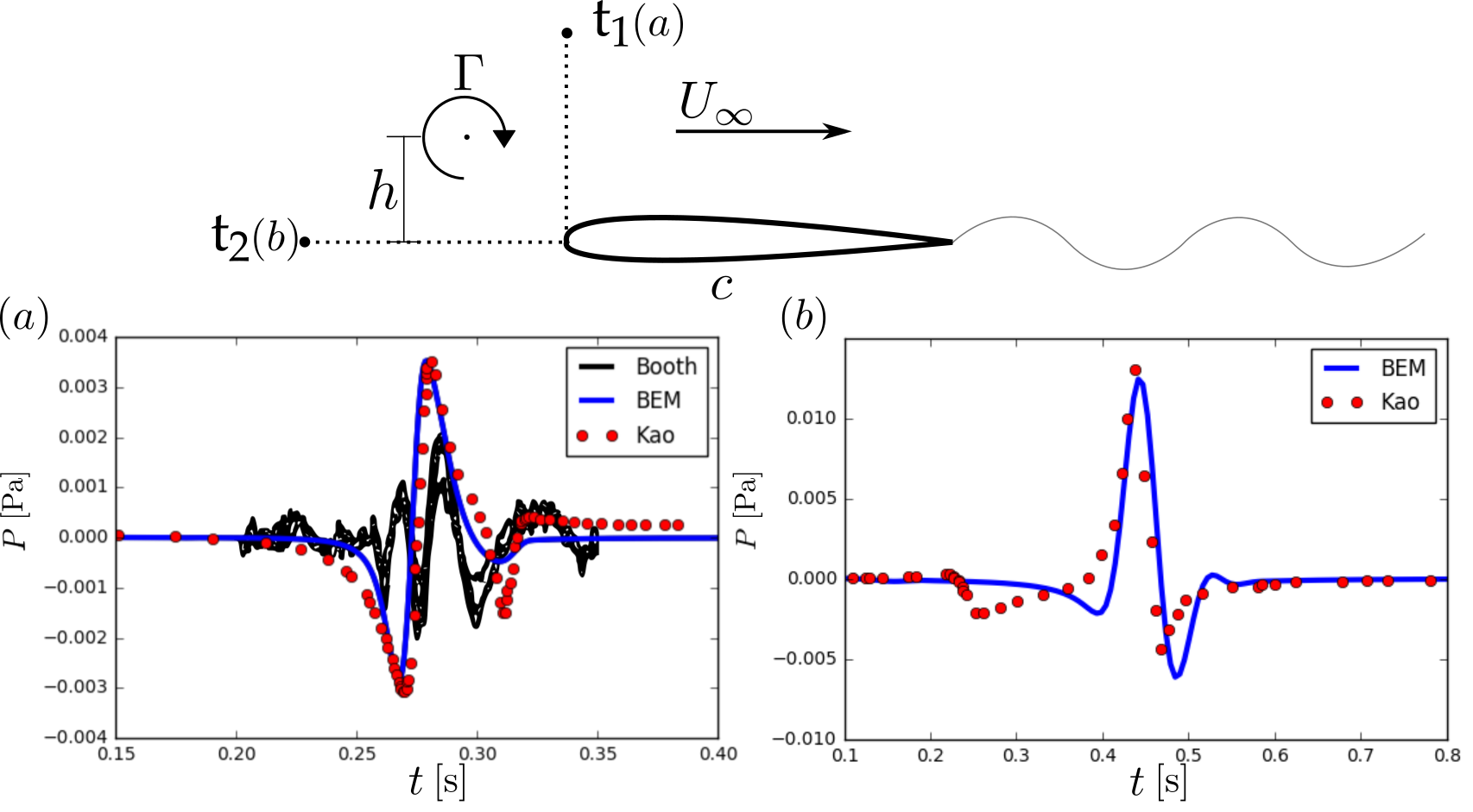}
    \caption{Acoustic emission due to a single vortex advecting past a NACA 0012 airfoil. The vortex circulation $\Gamma$, chord length $c$, observation point $\mathbf{t}$, freestream velocity $U_{\infty}$, and offset heights $h$ are different for each of the two validation cases. The response of $(a)$ is  observed 100 chord lengths in front of the foil $\mathbf{t}_1 = (-100c, 0)$, while the response of $(b)$ is observed 50 chord lengths above the airfoil $\mathbf{t}_2 = (0, 50c)$. The black lines represents the experimental results of Booth \cite{booth1990experimental}, the red circle represents the matched asymptotic solution of Kao \cite{kao2002body}, and the blue line is the result of the coupled potential flow and transient acoustics BEM put forward in this work.}
    \label{fig:vortex-body-validation}
\end{figure}

The experimental work of Booth  \cite{booth1990experimental} details acoustic scattering due to vortex-body interaction.   The matched asymptotic method of \citeauthor{kao2002body} \cite{kao2002body} uses this experimental study to validate their analysis. Kao asymptotically matches the acoustic loading from a potential flow solver to an outer far-field acoustic solution. The selected vortex-body interaction problem sets a vortex  upstream of a NACA 0012 airfoil, with a chord of $c = 0.2032$ m. The vortex has a circulation of $\Gamma = 0.52 $ m$^2$ s$^{-1}$ and advects in a freestream of speed $U_{\infty} = 4.7$ m s$^{-1}$ at a vertical displacement of $h = 0.152c$ from the foil centerline (cf.~Fig.~\ref{fig:vortex-body-validation}). This particular offset distance was selected because at other distances in the experimental study the vortex impinges on the body and breaks down. The fluid medium has a speed of sound of $c_0 = 343$ m s$^{-1}$ and the density $\rho = 1.225$ kg m$^{-3}$. The acoustic response is then found in front of the airfoil at $\mathbf{t}_1= (100c, 0)$, as shown in the problem schematic at the top of Fig.~\ref{fig:vortex-body-validation}.

Figure \ref{fig:vortex-body-validation}$(a)$ compares the experimental vortex-body interaction sound results from the  work of Booth, the matched asymptotic method of Kao, and the one way coupled potential flow and acoustic BEM presented in this study. The matched asymptotic method and the flow-acoustic BEM have qualitatively similar responses. The experimental acoustic response is the same order of magnitude as the other methodologies with similar qualitative trends, albeit with more fluctuations in its signal. The leading edge acoustic response occurs at $t \approx 0.25$ s. It can be seen that the amplitude of this interaction is quantitatively similar for the theoretical and BEM approaches, while qualitatively the slope of the leading-edge response as it approaches the minimum pressure from the flow-acoustic BEM is steeper than the matched asymptotic method response. 
The trailing-edge acoustic response occurs at $t \approx 0.30$ s, where the peak predicted response of $P\approx -0.0015$ Pa from the matched asymptotic solution is stronger than the BEM prediction of $P \approx -0.0005$ Pa. The increased pressure response at the trailing edge from the matched asymptotic method can be affected by the explicit Kutta condition applied and the manner in which the wake evolves behind the foil. Howe~\cite{howe_1976} stated that, as a vortex passes the trailing edge of a body, the vorticity shed into the wake tends to cancel the effect of the incoming vorticity and mitigates the noise generation. The Kutta condition in the potential flow BEM could be implicitly imposing the mechanism described by Howe, which would help explain the weaker acoustic response predicted by the flow-acoustic BEM framework. Additionally, the acoustic response in Fig.~\ref{fig:vortex-body-validation}$(a)$ is measured in front of the foil, a location where the acoustic pressure would be small in comparison to other measurement locations. 

%Next problem and similar results (b)
Further verification of the the flow-acoustic BEM is accomplished by measuring the acoustic response where it has its maximum value. Figure \ref{fig:vortex-body-validation}$(b)$ presents the measurement of the acoustic response above the foil where the peak acoustic pressure occurs. The flow scenario has a vortex with circulation $\Gamma =$ 0.1 m$^2$ s$^{-1}$ that is released five chord lengths upstream with  vertical offset $h = 0.1c$  above the center of a NACA 0012 foil with chord $c = 1$ m. The case has a freestream velocity $U_{\infty} = 1$ m s$^{-1}$, a sound speed of $c_0 = 5$ m s$^{-1}$, and density $\rho$ = 1 kg m$^{-3}$.  The acoustic response was measured fifty chord lengths above the leading edge of the foil at $\mathbf{t}_2 = (0,50c)$.  The matched asymptotic method predicts a leading-edge acoustic response  at $t\approx 0.25$ s, which occurs before the flow-acoustic BEM response at $t\approx 0.35$ s. However, both approaches exhibit a similar magnitude of response. The maximum acoustic responses of both systems are found at  $t\approx 0.45$ s with a pressure of $P = 0.013$ Pa.  The trailing-edge responses at $t\approx 0.5$ s also exhibits similar magnitudes for both approaches. 
%the pass
The flow-acoustic BEM predicts the  same order-of-magnitude acoustic response as the experiments of \citeauthor{booth1990experimental} \cite{booth1990experimental}. In addition, the flow-acoustic BEM produces qualitatively and quantitatively similar acoustic responses to the matched asymptotic method of Kao \cite{kao2002body}.

\section{Acoustic Emission from Biological Swimming}\label{S:bio-swimming}
\begin{figure}
\centering
\includegraphics[width=0.55\textwidth]{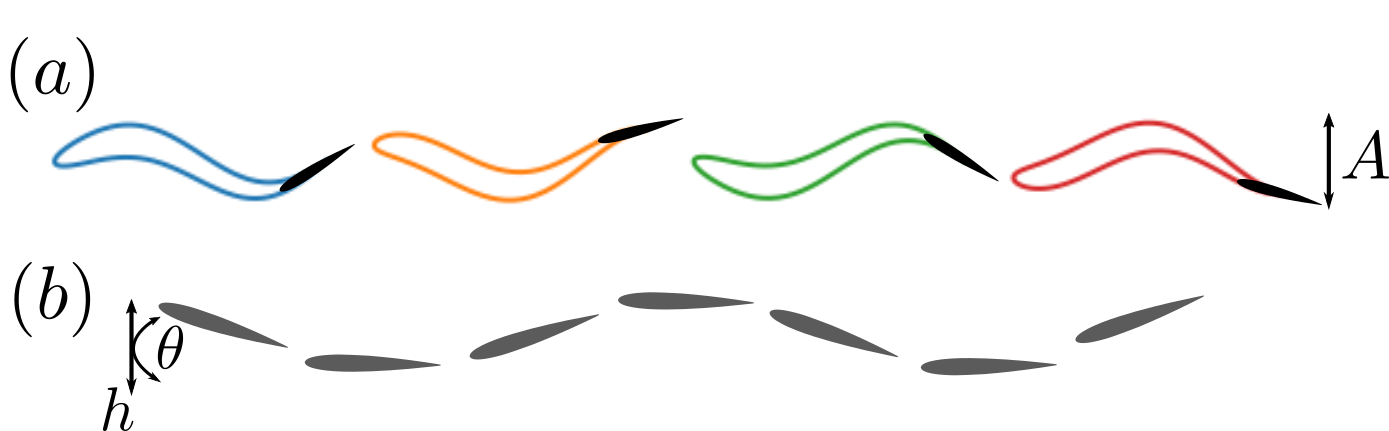}
\caption{Use of a pitching/heaving foil as a proxy to undulatory locomotion. Illustration $(a)$ shows a period of a traveling wave undulating across a NACA 0012 airfoil. The trailing edge of the foil is modeled as a separate entity that acts as a proxy to the caudal fin of a fish. The schematic $(b)$ tracks the motion of the `caudal fin' separated from the body as a function of pitching and heaving.
}
\label{fig:foil_schematic}
\end{figure}

Many fish swim by undulating their bodies and oscillating their caudal or tail fins, which are in many cases responsible for the majority of their thrust production \cite{sfakiotakis1999review,lighthill1969hydromechanics}.  A common and simple representation of a biological swimmer is to neglect the body and only consider the caudal fin as a combined heaving and pitching hydrofoil \cite{akoz2018unsteady}, which is also the case in the present study. Fish-like locomotion via a traveling wave is shown in Fig.~\ref{fig:foil_schematic}$(a)$. The motion of the rear of the foil is tracked and then treated as a discrete propulsive foil, which is denoted in the figure as a solid black body. Figure \ref{fig:foil_schematic}$(b)$ shows how a combined heaving and pitching motion of a rigid body is used as a proxy for the entire traveling wave system. The rigid body pitches about its leading edge, as that is where it would be connected to the fish.

The heaving and  pitching motion and the peak-to-peak amplitude of the foil are described by
\begin{align}
h(t) = h_0 \sin(2\pi ft),\\
\theta(t) = \theta_0 \sin(2\pi ft + \phi),\\
A(t^*) = 2 \, \text{max} \{h(t) + c \, \sin\left[\theta(t)\right]\}, \\
h^* = 2\, h(t^*) / A(t^*), \\
\theta^* = 2\, c \, \sin\left[\theta(t^*)\right] / A(t^*), \\
A^* = A(t^*)/c,
\end{align}
\noindent where $f$ is the frequency of motion, $t$ is time, and $h_0$ and $\theta_0$ are the maximum heaving amplitude and maximum pitching angle, respectively. The phase delay between heave and pitch signals is $\phi = \pi/2$. The peak-to-peak amplitude is $A(t^*)$, and $t^*$ is the time at which the foil reaches its peak amplitude. Given $A(t^*)$ and $\theta_0$, $h_0$ can be calculated by a nonlinear equation solver. Normalizing $h(t^*)$ and $\theta(t^*)$ by $A(t^*)$ produces the identity
\begin{equation}
h^* + \theta^* = 1.
\end{equation}
%%define ND variables

Supposing a non-dimensional peak-to-peak amplitude $A^*$, the ratio of the heaving and pitching amplitudes is solely described by the non-dimensional heave-to-pitch ratio $h^*$. A purely pitching foil has a value of $h^*=0$, a purely heaving foil has a value of $h^*=1$, and $h^* =0.5$ represents a combined heaving and pitching motion where half of the total amplitude comes from pitching and the other half comes from heaving.  All other values in the range $0 < h^* < 1$ represent combined heaving and pitching motions.  The chord-based reduced frequency $f^* =fc/U_{\infty}$ is the non-dimensional quantity that describes the unsteadiness of the prescribed swimming motion. 

\begin{figure}
\centering
\includegraphics[width=0.7\textwidth]{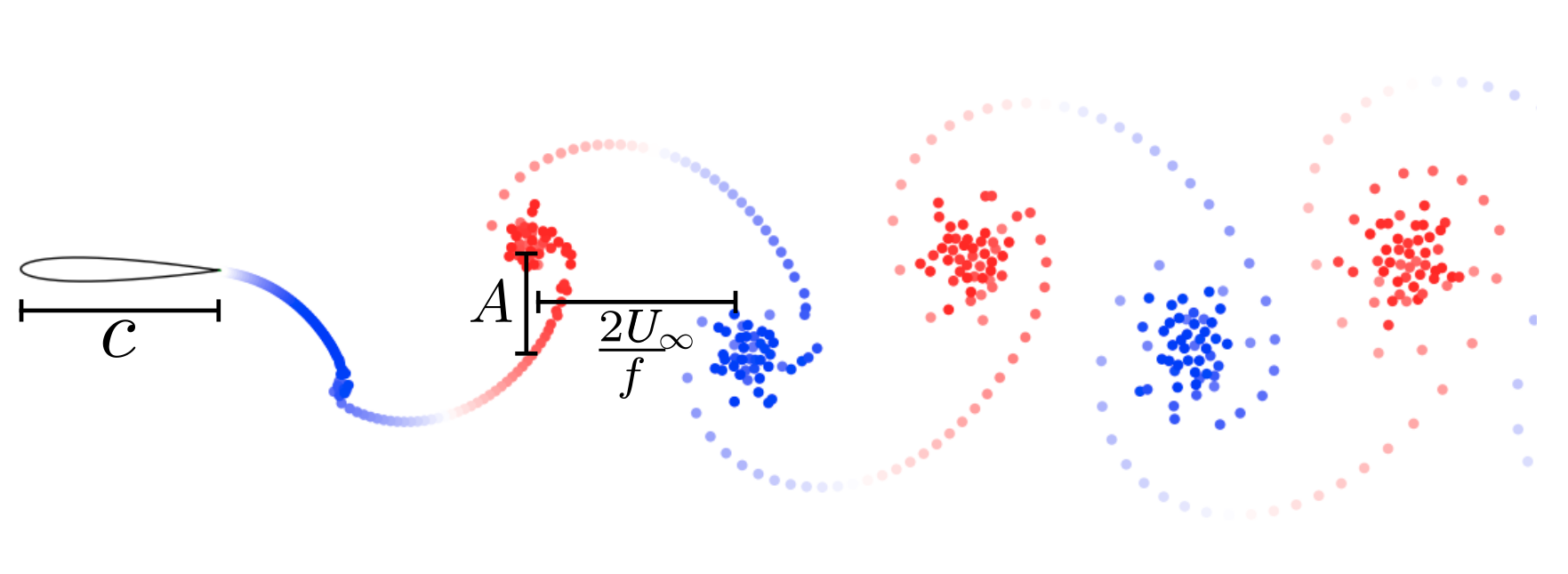}
\caption{Typical wake of an unsteady swimmer in this study. Wake of a foil after several cycles of motion for the values of $h^* = f^* =  A^* = 0.5$. A foil of chord $c$ is placed in a freestream flow at speed $U_{\infty}$. The vertical spacing of the vortices in the wake are described as a function of the amplitude $A = c/2 $ and the horizontal spacing is a function of freestream speed and frequency, $2U_{\infty}/f$, for half of a cycle of motion.}
\label{fig:a05h05f05wake}
\end{figure}

Figure \ref{fig:a05h05f05wake} shows a typical reverse \vK wake structure of a foil operating at $h^* =0.5$, $f^* = 0.5$, and $A^*= 0.5$. The spacing of the wake is dictated vertically by the amplitude of motion  and horizontally by the freestream speed and the frequency. The ratio of the vertical spacing to horizontal spacing results in the Strouhal number, $St = fA(t^*)/U_{\infty}$. The example wake in Fig.~\ref{fig:a05h05f05wake} has a Strouhal number of $St = 0.25,$ which is within the range of typical fish swimming of $0.2 < St < 0.4$ \cite{triantafyllou1995efficient,taylor2003flying,eloy2012optimal}. The acoustic pressure presented for the remainder of this work is non-dimensionalized by dynamic pressure, $P^* = 2P_{\rm{acoustic}}/\rho U_{\infty}^2.$

%input table
The flow-acoustic BEM is used to study how variations in the non-dimensional amplitude, reduced frequency, and non-dimensional heave-to-pitch ratio  alter both the acoustic emissions and the hydrodynamic forces on the body. The ranges of variables and parameters used in the current study are presented in Table \ref{tab:inputs}.  The range of Strouhal numbers in the simulations is $0.0312 \le St \le 1$. The reduced frequencies and Strouhal numbers in the current study cover the ranges associated with typical fish swimming \cite{eloy2012optimal,triantafyllou1995efficient} and also extend to regions where biological systems may perform fast starts or rapid turns \cite{sfakiotakis1999review}. 

\begin{table}
 \begin{center}
  \begin{tabular}{lccc}
  \hline
    \textbf{Input Variables:} & $0.25 \leq A^* \leq 1$ & $0 \leq h^* \leq 1$  \\ &  $0.125 \leq f^* \leq 1$ &  &\\
    \textbf{Input Parameters:} & $U_{\infty} = 1$ m s$^{-1}$ & ~$\rho = 1000$ kg m$^{-3}$ & 
    \\
    & ~~$c_0 = 1000$ m s$^{-1}$ &  ~$c = 1$ m &  \\ 
    \hline
  \end{tabular}
  \caption{Input variables and parameters used in the present study.}{\label{tab:inputs} }
 \end{center}
\end{table}

\section{Results}\label{S:results}
\begin{figure}
\centering
\includegraphics[width=12.9cm]{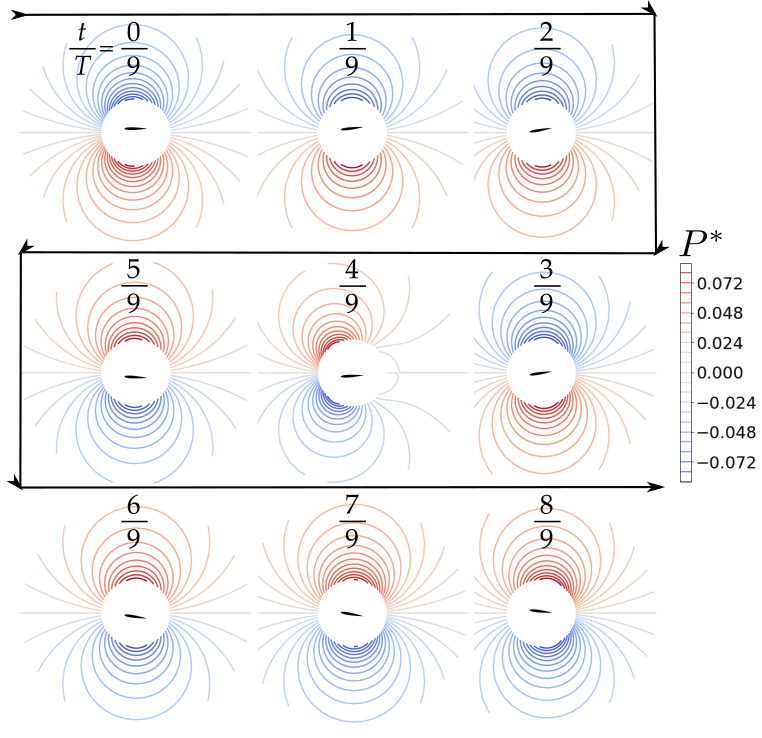}
\caption{Transient near-field acoustic pressure of a typical swimmer. The non-dimensional near-field acoustic pressure $P$ is shown for a foil operating at $h^* = 0.5$, $f^* = 0.5$, and $A^* = 0.5$ at different instances of non-dimensional time $t/T$.  The acoustic pressure is found at discrete points set around circles centered about the mid-chord when at rest. The circle radii range from two to five chord lengths. }
\label{fig:transMap}
\end{figure}

%Define transient acoustic field
Figure \ref{fig:transMap} presents the near-field transient acoustic pressure for a foil with parameters of $h^* = 0.5$, $f^* = 0.5$, and $A^* = 0.5$.  The acoustic pressure is determined at discrete points on circles with radii of two to five chords away from the mid-chord of the foil at rest. A vertically-oriented pressure dipole is generally observed. The position of the maximum acoustic pressure shifts from the front to back as the effective angle of attack increases, as seen in the snapshots from $t/T = 0$ to $t/T = 5/9$, where $T$ is the period of motion. The sign change of the dipole strength at the middle of the period ($t/T = 4/9$) corresponds to the change in effective angle of attack going from negative to positive values. At time $t/T = 4/9$, the transient acoustic pressure field has a quadrupole shape with  two of its lobes directed behind the foil, which are an order of magnitude weaker than the response above and below the foil. 

\begin{figure}
\centering
\includegraphics[width=12.9cm]{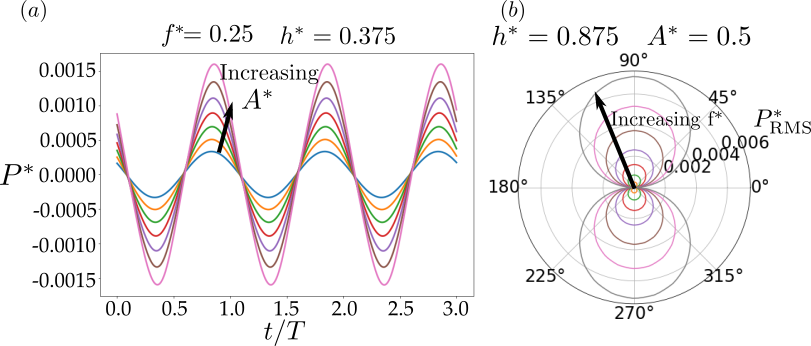}
\caption{Transient and root-mean-square (RMS) acoustic pressure levels of a typical swimmer. $(a)$ Transient acoustic pressure response of a foil undergoing a combined heaving/pitching motion with $h^* = 0.375$, $f^* = 0.25$, and various amplitudes. The acoustic pressure is determined at the position 50 chords above the leading-edge of the foil and shown for 3 cycles of motion.  $(b)$ RMS acoustic pressure $P_\text{RMS}$ from a foil undergoing a combined heaving/pitching motion with $h^* = 0.8$, $A^*=0.5$, and various reduced frequencies. The acoustic pressure response is computed on a circle 50 chord lengths from the leading edge of the foil and is averaged over 3 cycles of motion. The dipolar acoustic directivity is observed for all kinematic parameters considered.}
\label{fig:char_trans_rms}
\end{figure}

%%Define Acoustic measurements -- why peak RMS
Figure \ref{fig:char_trans_rms}a shows the acoustic response at a single observation point 50 chords above the foil with a heave-to-pitch ratio of $h^* = 0.375$, a reduced frequency of $f^*=0.25$, and over the entire range of amplitudes used in the current study.  As expected, the pressure fluctuates harmonically in time with the same frequency as the foil motion.  For fixed heave-to-pitch ratio and reduced frequency, the amplitude of the acoustic pressure response increases with the amplitude of motion. A directivity plot of the root-mean-square (RMS) pressure ($P_\text{RMS}$) over three motion cycles for various reduced frequencies is shown in Fig.~\ref{fig:char_trans_rms}$(b)$. Regardless of the motion parameters selected, a dipole directivity is always observed.  Moreover, the peak acoustic pressure can be observed to also increase with an increase in the reduced frequency.  Since all of the variables produce self-similar dipole acoustic responses, the peak RMS acoustic pressure can be used as the single metric to describe the acoustic field. The peak pressure can then be scaled by the dynamic pressure:
\begin{equation}
P^*_{\rm{Peak}} = \frac{P^*_\text{RMS}}{\frac{1}{2}\rho U^2}.
\end{equation}

\begin{figure}
\centering
\includegraphics[width=12.9cm]{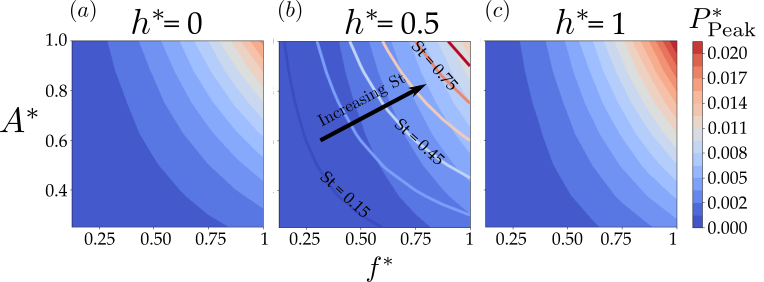}
\caption{Peak acoustic pressure for $(a)$ a purely pitching foil  ($h^* = 0$), $(b)$ a combined heaving/pitching foil ($h^* = 0.5$), and $(c)$ a purely heaving foil ($h^* = 1$) as a function of reduced frequency and amplitude of motion. Isolines of Strouhal number overlay the acoustic pressure contours in $(b)$.}
\label{fig:heave_study}
\end{figure}

%%Effect of mixed motion on acoustic output
Figure \ref{fig:heave_study} presents the non-dimensional peak acoustic pressure for a purely pitching foil ($h^* = 0$), a combined-motion foil with equal amplitude contributions from pitching and heaving ($h^* = 0.5$), and a purely heaving foil ($h^* = 1$). The global maximum in the peak acoustic pressure is found at $h^*=1$, $f^*=1$, and $A^*=1$, which represents the upper bounds of all of the parameters being explored. However, the minimum in the noise production does not occur at the minima of the parameter set: the minimum noise occurs at $h^*\approx 0.25$ for fixed values of amplitude and reduced frequency.  A purely heaving foil produces higher acoustic pressures than a purely pitching foil, except for  $f^* \lesssim 0.25$. The combined heaving and pitching foil emits a weaker acoustic pressure signal than either purely heaving or pitching foils for all combinations of reduced frequency and amplitude of motion. Figure \ref{fig:heave_study}$(b)$ overlays the Strouhal number on the peak acoustic pressure, showing that in general an increase in Strouhal number results in an increase in the peak RMS acoustic pressure for a fixed swimming motion $h^*$, even though the isolines of $St$ and $P^*_{\rm{peak}}$ are not precisely aligned. Therefore, the noise level trends are not driven solely by changes in the Strouhal number.

\begin{figure}
\centering
\includegraphics[width=8.6cm]{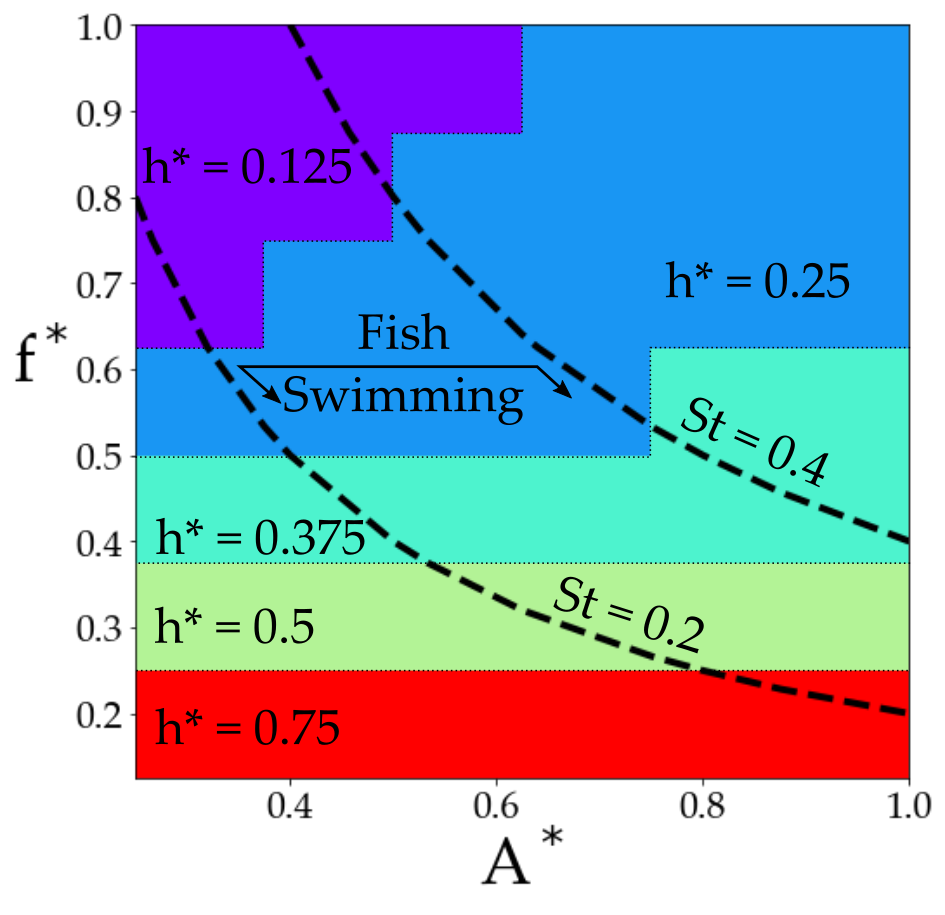}
\caption{Map showing for a given $f^*$ and $A^*$ which $h^*$ value leads to the lowest noise production.}
\label{fig:heave_mins}
\end{figure}

%%Why? %% where of mixed motion on acoustic output
An increase in either the reduced frequency or the amplitude of motion will result in an increase in acoustic pressure, but the motion type of the foil can lead to lower values of acoustic pressure. In fact, a pressure minimum is observed for a combined heaving and pitching motion for all combinations of reduced frequency and amplitude examined in this study. Figure \ref{fig:heave_mins} presents a map that shows the $h^*$ value leading to the lowest noise production for a given $f^*$ and $A^*$.  A purely heaving or pitching foil never produces the lowest acoustic pressure.  For $f^* \gtrsim 0.3$ the quietest swimming is produced by pitch-dominated swimming motions ($h^* < 0.5$).  Only for low reduced frequencies of $f^* \lesssim 0.3$ do heave-dominated motions ($h^* > 0.5$) produce the quietest swimming, and this $h^*$ value is independent of the amplitude.  Lines of constant $St$ are also marked on the figure, which denote the typical range of $St$ for swimming animals \cite{taylor2003flying}.  From this map it can be seen that for swimming animals  operating with low reduced frequencies ($f^* < 0.3$) their noise production is minimized if they utilize heave-dominated swimming kinematics. Swimming animals with higher  reduced frequency ($f^*>0.3$) minimize their noise production if they utilize pitch-dominated swimming kinematics.

%outputs - hydrodynamic
The potential flow solver also can solve for the associated performance characteristics of these oscillating foils. The force acting on the foil is defined by $\mathbf{F} = \int_{S_{\rm{b}}}-(P_\text{flow} \hat{\mathbf{n}})\, dS$, where $P_\text{flow}$ is the pressure from the flow solver as opposed to the acoustic pressure, $\hat{\mathbf{n}}$ is the local outward normal vector and $S_b$ is the foil surface.  Since the potential flow method is inviscid, the forces on the foil arise only from its external pressure distribution.  The power consumption of the oscillating motion is calculated as the negative inner product of the force vector and velocity vector of each boundary element, i.e., $P_w = -\int_{S_b} \mathbf{F}_{\rm{ele}} \cdot \mathbf{u}_{\rm{ele}} \, dS$.  The time-averaged coefficients of lift, thrust, and power may be defined as
\begin{eqnarray}
C_L = \frac{\overline{F_y}}{\frac{1}{2}\rho U_{\infty}^2},\quad 
C_T = - \frac{\overline{F_x}}{\frac{1}{2}\rho U_{\infty}^2}, \quad
C_P =  \frac{\overline{P_w}}{\frac{1}{2}\rho U_{\infty}^3}, 
\end{eqnarray}
where $F_x$ and $F_y$ are the integrated streamwise and transverse components of the force on the foil, respectively. The efficiency is also defined as $\eta = {C_T}/{C_P}$. In addition to the time-averaged coefficient of lift, the maximum coefficient of lift will also prove to be a useful metric and is defined by 
\begin{equation}
C_{L,\text{max}} = \frac{{\rm max}(\bar{F}_y)}{\frac{1}{2}\rho U_{\infty}^2}. 
\end{equation}

\begin{figure}
\centering
\includegraphics[width=12.9cm]{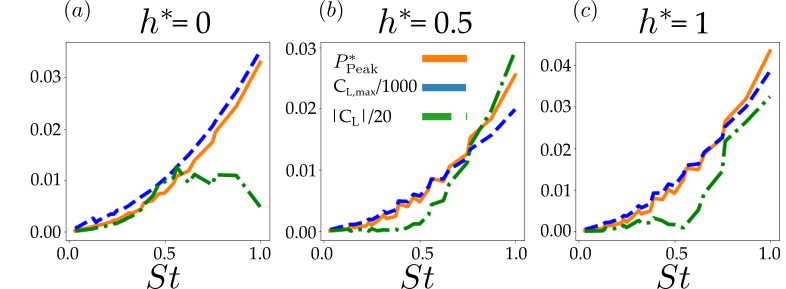}
\caption{Comparison of lift metrics with peak acoustic pressure as a function of $St$ and $h^*$. The plots show the peak RMS acoustic pressure $P^*_{\rm{Peak}}$, maximum coefficient of lift scaled $C_{L\text{max}}/1000$, and the absolute value of the time-averaged coefficient of lift $|C_L|$ scaled by a factor of $20$.  Subfigure $(a)$ is a purely pitching foil, $h^* = 0$, $(b)$ is a combined heaving and pitching motion with $h^* = 0.5$, and $(c)$ is a purely heaving foil, $h^* = 1$.}
\label{fig:Pave_clmax}
\end{figure}

Figure \ref{fig:Pave_clmax} presents a comparison of the peak RMS acoustic pressure $P^*_{\rm{Peak}}$, the maximum coefficient of lift $C_{L,\text{max}}$ scaled by a factor of $1000$ and the absolute value of the time-averaged coefficient of lift $|C_L|$ scaled by a factor of $20$.  These quantities are shown as functions of the Strouhal number and non-dimensional heave-to-pitch ratio. The lift metrics were scaled in order to make the plots the same order of magnitude. It can be seen that the peak RMS acoustic pressure and the maximum coefficient of lift follow the same trend for increasing $St$ and $h^*$. It can also be observed that the peak RMS acoustic pressure does not follow the same trend as the time-averaged coefficient of lift. 
Not surprisingly, the maximum lift coefficient provides a better correlation with the peak RMS acoustic pressure than the time-average coefficient of lift. In light of this finding, the maximum coefficient of lift $C_{L,\text{max}}$  may be used as a proxy metric for comparison of the relative acoustic emissions between oscillating hydrofoils.

%Switch to lines plots of P,Cl,Clmax vs St- makes a better argument

\begin{figure}
\centering
\includegraphics[width=1\textwidth]{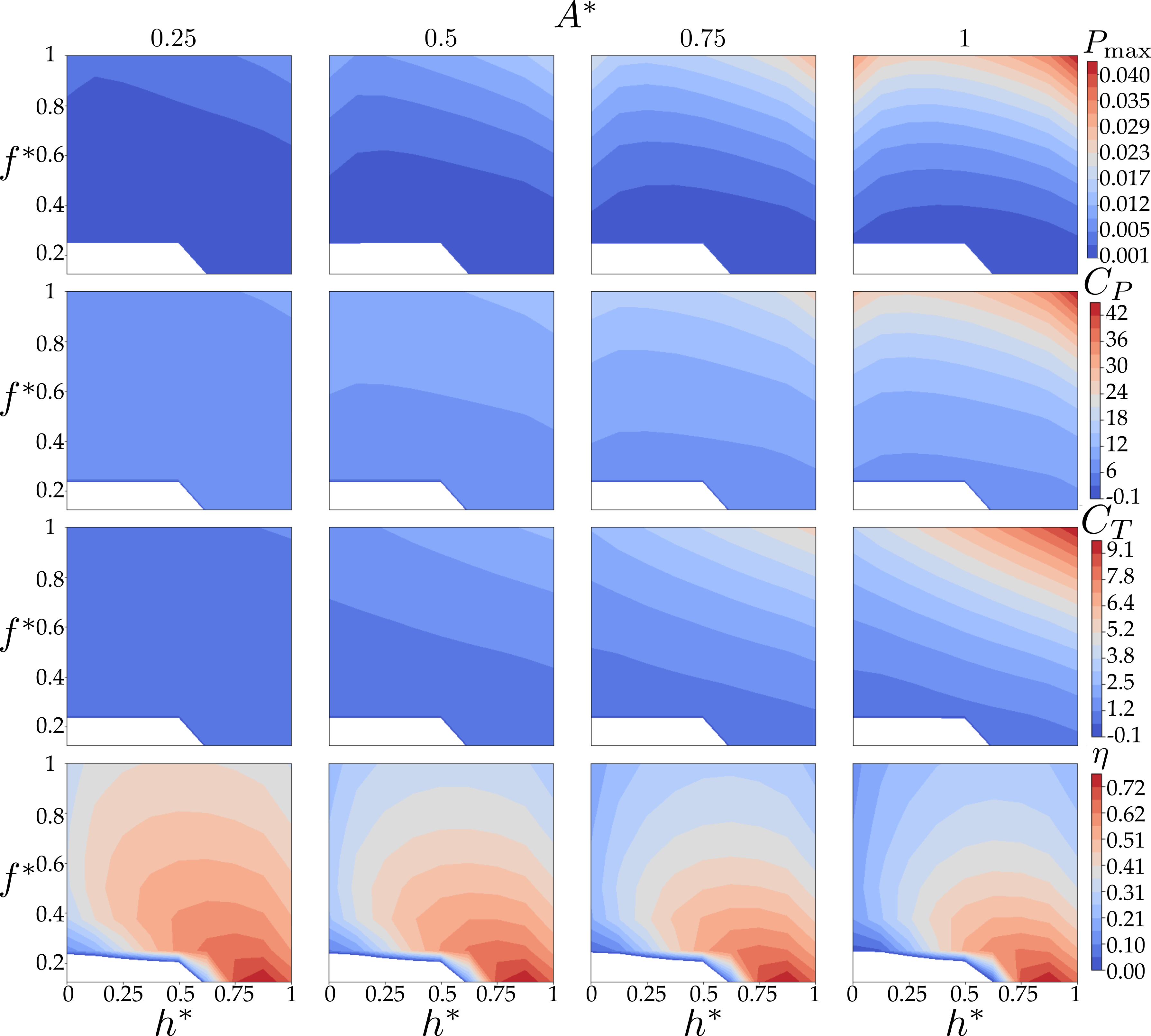}
\caption{Comparison of acoustic and hydrodynamic metrics. The rows correspond to the coefficients of thrust and power, the efficiency, and the peak acoustic pressure.  The columns correspond to different $A^*$ values.  Each contour plot is presented as a function of $f^*$ and $h^*$.}
\label{fig:maxPerfComp}
\end{figure}

%Big data excavation
Figure \ref{fig:maxPerfComp} presents a comparison among the coefficients of thrust and power, efficiency, and the peak RMS acoustic pressure.  The thrust production increases with increasing $f^*$, $h^*$, and $A^*$. In figure  \ref{fig:maxPerfComp}, regions of negative coefficient of thrust (and the corresponding regions of coefficients of power, efficiency, and the peak RMS acoustic pressure) are excluded from the contour plots shown. The power consumption also increases with increasing $f^*$ and $A^*$; however, as $h^*$ increases the power decreases to a minimum and then increases.  This result indicates that combined heaving and pitching motions use less power than purely pitching or heaving motions for a fixed $f^*$ and $A^*$.  The efficiency results show global peaks around $h^* = 0.85$, $f^* < 0.2$ for all $A^*$.  Since the $St = f^*A^*$, the highest efficiencies occur for the lowest swimming Strouhal numbers.  Most fish swim with $0.2 \leq St \leq 0.4$ \cite{taylor2003flying} making it difficult to reach the highest levels of efficiency, even with $A^* = 1$.  The optimal $h^*$ to maximize efficiency will vary, depending upon the Strouhal number of the particular swimming animal or biorobotic device.

For the first time,  the noise of a biopropulsor and its performance can be compared.  The peak acoustic pressure can be observed to not follow trends of the thrust or efficiency; however, the peak acoustic pressure does follow similar trends with the power coefficient.  This result can be explained by the fact that the acoustic pressure is well-correlated with the maximum lift coefficient and, consequently, with the power consumption \cite{moored2018inviscid}.  Moreover, the absolute value of coefficient of lift is greater than thrust for all scenarios, further explaining the vertical acoustic dipole and the trend between the peak RMS acoustic pressure and the maximum coefficient of lift.  

These results highlight that when the power needed to move the foil is minimized, then there is also a minimum amount of energy that can be converted into noise, leading to the quietest acoustic signatures. In contrast, there is a trade-off between operating at maximum propulsive efficiency and minimizing the noise production.  Furthermore, the thrust increases as the swimmer moves from a pure pitching to a pure heaving swimming motion, which requires more power and produces a louder acoustic signal. 

\section{Conclusion}\label{s:conclusion}
An integrated, two-dimensional flow-acoustic boundary element solver is developed to predict the noise generated by the vortical wake of rigid foils in motion. The vortex-particle wake computed by the potential flow boundary element solver furnishes the input for the transient acoustic boundary element solver via Powell's acoustic analogy. This one-way flow-acoustic coupling is validated against experimental and analytical results for the acoustic emission of a vortex gust encounter with an airfoil.

The coupled potential flow-acoustic method is used to investigate the performance and acoustic emission of a heaving and pitching hydrofoil. The hydrofoil is subjected to varying non-dimensional frequencies, amplitudes, and heave-to-pitch ratios that encompass the parametric range of most swimming and maneuvering animals. All combinations of these variables examined in this work produce a similar dipole acoustic response, where the maximum sound pressure levels occur directly above and below the foil. Foils in purely pitching or purely heaving motions are found to be noisier than foils that operate with a combined heaving and pitching motion.  In fact, for fixed reduced frequency and amplitude there exists an optimal heave-to-pitch ratio, $h^*$, that minimizes the noise production.  The numerical model indicates that most swimming animals would minimize their noise production by using heave-dominated swimming motions. As the reduced frequency increases past the regime of swimming animals, a transition to pitch-dominated swimming motions minimize the noise production.  Moreover, the correlation between the maximum coefficient of lift and the peak RMS acoustic pressure for all combinations of reduced frequency, amplitude, and heave-to-pitch ratio corresponds to the acoustic dipole response. Consequently, the trends in the coefficient of power are well-correlated with the trends in the peak RMS acoustic pressure for swimming motions with $h^* > 0.25$.  This result supports the conclusion that swimming with low power consumption and a low acoustic signature can be achieved together.  In contrast, it is discovered that there is a trade-off between swimming with high propulsive efficiency and a low acoustic signature.  These insights seek to further our understanding of swimming in nature and could aid in the design of high-performance, quiet bio-inspired autonomous underwater vehicles. 

\section*{Acknowledgments}
The authors gratefully acknowledge financial support from the National Science Foundation under grants 1805692 (JWJ) and 1653181 (KWM), the Office of Naval Research under MURI grant N00014-08-1-0642 (KWM), and a Lehigh CORE grant (JWJ, KWM).

\appendix
\section{}\label{s:app_potFlow}
The potential flow boundary element method presented in this work is a two-dimensional derivative of the three-dimensional method described by Willis \textit{et al.}~\cite{willis2007combined}. A comparison of the BEM solution to analytic and numerical works is performed to ensure accuracy of the method presented. Theodorsen~\cite{theodorsen1935general} solved analytically for the fluid-dynamic lift and moment acting on a flat-plate foil undergoing harmonic pitching and heaving motions under the assumption of a planar wake. Garrick~\cite{garrick1937propulsion} used these results to predict the time-averaged thrust and efficiency of pitching and heaving motions, as well as trailing-edge flap motions. These analytical results have recently been extended by Jaworski~\cite{jaworski2012thrust} to also handle leading-edge flap actuation in addition to pitch, heave, and trailing-edge flap motions, and the results from Theodorsen and Garrick have previously been compared to computational fluid dynamic simulations of rigid and deformable thin airfoils~\cite{jaworski2012high,jaworski2015thrust}.

%A few years later, Garrick extended the formulations of Theodorsen to compute forces for  heaving and/or pitching foils\cite{garrick1937propulsion}.

The first validation case is against Garrick's theory. In the numerical simulations, a 2~\%-thick tear drop foil is subjected to a purely pitching motion of amplitude $\theta_0 = 3^{\circ}$ about the leading edge. 

\begin{figure}
    \centering
    \includegraphics[width=0.95\textwidth]{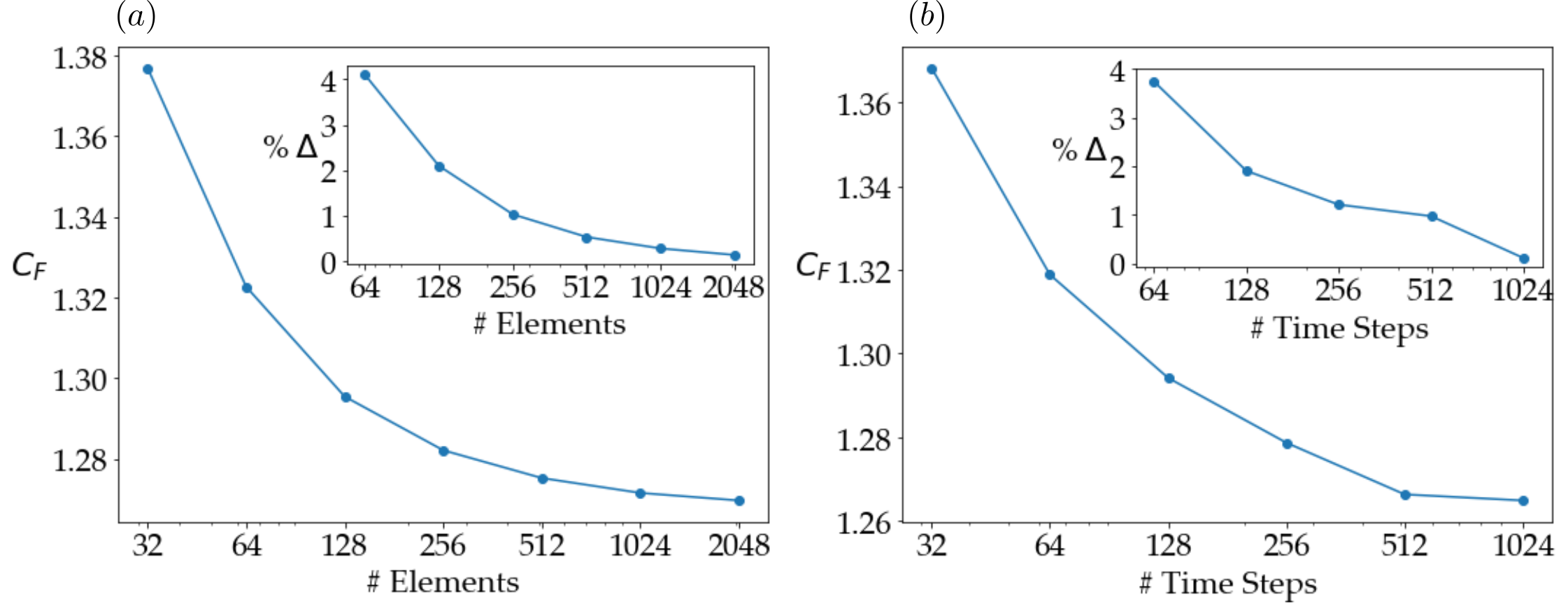}
    \caption{Spatial and temporal convergence of potential flow BEM. (a) Plots the time averaged coefficient of force as the number of boundary elements on the body doubles for a fixed number of time steps. }
    \label{fig:convergence}
\end{figure}

First, convergence studies on the number of boundary elements and time-steps is found for a reduced frequency $f^* = 1$. Figure \ref{fig:convergence} $(a)$ shows spatial and $(b)$ temporal convergence for time-averaged coefficient of force. The inset images detail the   percent change (\%$\Delta$) of the time-averaged value as the number of elements or time steps per period of motion double. The spatial convergence was conducted for a fixed temporal resolution of 150 time steps per period. The inset of figure \ref{fig:convergence} $(a)$ shows the ${\it O}(1\%)$ difference in the force when changing  from 128 to 256 elements. The temporal convergence study used a fix number of 256 body elements. It can be seen in figure \ref{fig:convergence} $(b)$  an ${\it O}(1\%)$ change in force when increasing from 128 to 256 time steps per period of motion. All of the simulation results previously presented used 150 time steps per period of motion and 256 boundary elements to define the discrete body. Comparison to analytic solutions and convergence studies of the acoustic BEM are presented in Appendix \ref{s:plane_wave}. The acoustic results have been previously conducted by \citeauthor{wagenhoffer2017accelerated} \cite{wagenhoffer2017accelerated}, showing the selected spatial and temporal values of the potential flow solver will also result in converged acoustic results. 

Next, validation via comparison of Garrick's theory to the solution of the solver in \S \ref{S:flow_bem} for varying reduced frequencies is detailed. 
Theodorsen defined lift as
\begin{equation}
    C_L = \rho V^2 c \sqrt{R^2+I^2}e^{\i\omega t},
\end{equation}
and aerodynamic moment as, 
\begin{equation}
    M = \frac{1}{2}\rho V^2 c^2 \sqrt{R^2+I^2}e^{\i(\omega t + \phi)},
\end{equation}
where
\begin{eqnarray*}
    R = \pi \theta_0\left\{\frac{k^2}{2}\left(\frac{1}{8}+a^2\right)+\left(\frac{1}{2}+a\right)\left[F - \left(\frac{1}{2}-a\right)kG\right]\right\},\\
    I = -\pi \theta_0\left\{\frac{k}{2}\left(\frac{1}{2}-a\right)-\left(\frac{1}{2}+a\right)\left[G + \left(\frac{1}{2}-a\right)kF\right]\right\}, \\
    \phi = \tan^{-1}\frac{I}{R}.
\end{eqnarray*}
The required power to sustain the foil motion is ${\rm Pow} = -M \dot{\theta}.$
The coefficient of thrust from Garrick \cite{garrick1937propulsion} was corrected by Jones \textit{et al.}~\cite{jones1996wake} to be
\begin{equation}\label{eqn:Garrick}
    C_T = \pi k^2  \theta_0^2 \left[ (F^2 + G^2)\left(\frac{1}{k^2} + \left(\frac{1}{2}-a\right)^2\right) - \left(\frac{1}{2}-a\right)\left(\frac{1}{2} - F\right) - \frac{F}{k^2} -\left(\frac{1}{2}+a\right)\frac{G}{k}\right],
\end{equation}
where $F$ and $G$ are the real and imaginary parts of the Theodorsen lift deficiency function, $C(k) = \i H_1^{(1)}(k)/(H_0^{(1)}(k)+\i H_1^{(1)}(k))$ and $a$ is the position of the pivot point measured from the mid-chord in half-chord intervals. Rotation about the leading edge corresponds to $a=-1$.

\begin{figure}
    \centering
    \includegraphics[width=12.9cm]{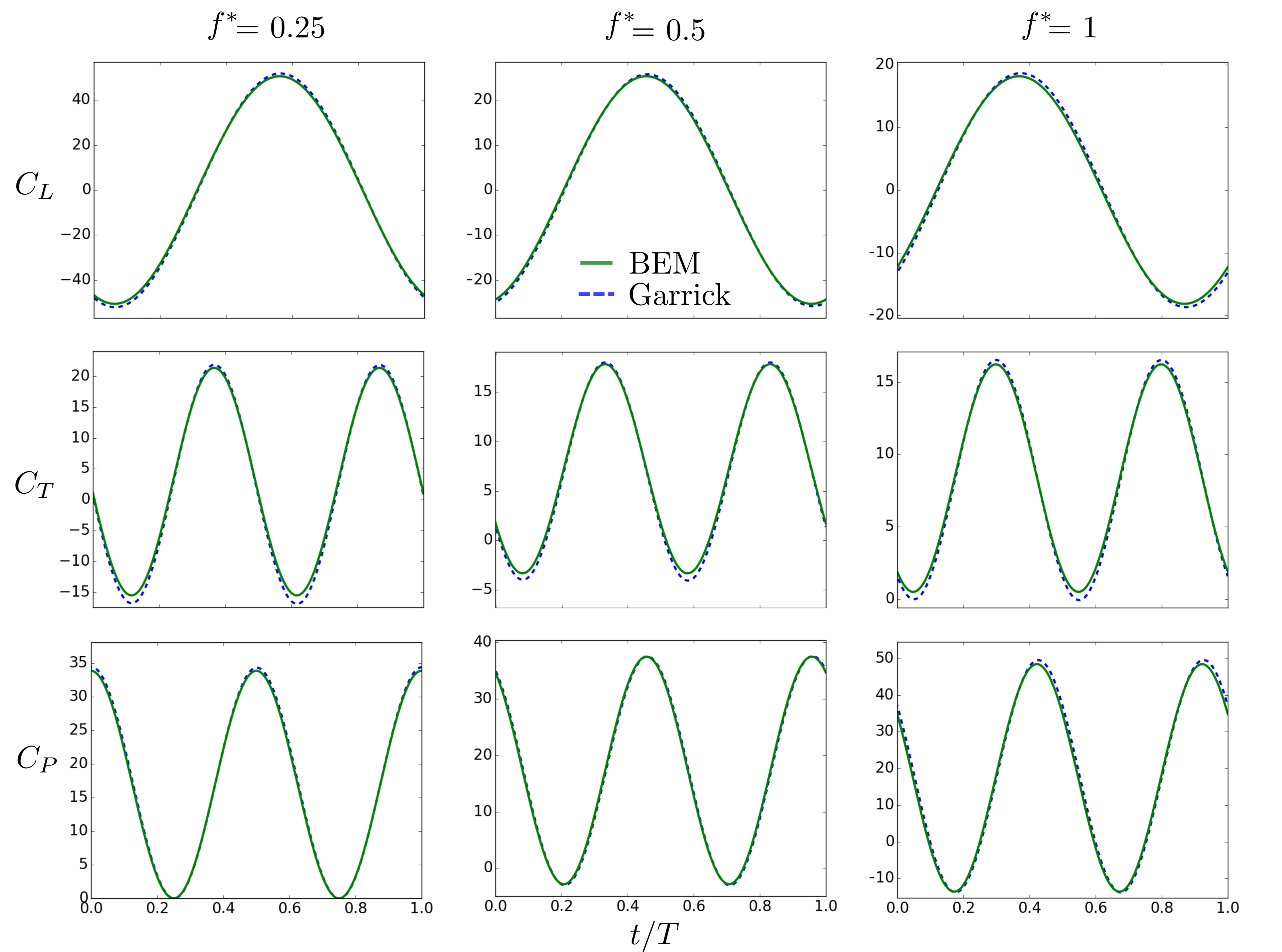}
    \caption{Comparison of BEM solution with solution of Garrick. Shown is a comparison of the solution of \S\ref{S:flow_bem} to the theory of Garrick \cite{garrick1937propulsion} for a purely pitching foil. From left to right are increasing values of reduced frequency, $f^* = fc/U_{\infty} = [0.25, 0.5, 1]$. From top to bottom are comparisons of the coefficient of lift $C_L$, coefficient of thrust $C_T$, and coefficient of power $C_P$, respectively. In each plot the dashed blue line is the solution of the potential flow method and the solid green line is the theory of Garrick \ref{eqn:Garrick}.}
    \label{fig:garrick_comp}
\end{figure}

Figure~\ref{fig:garrick_comp} compares the potential flow BEM results against the theory of Garrick for three different reduced frequencies that encapsulate the range of reduced frequencies used for the study in \S\ref{S:bio-swimming}. The figure shows the BEM solution matching well to the thin-airfoil theory results over one cycle of motion. 

\begin{figure}
    \centering
    \includegraphics[width=8.6cm]{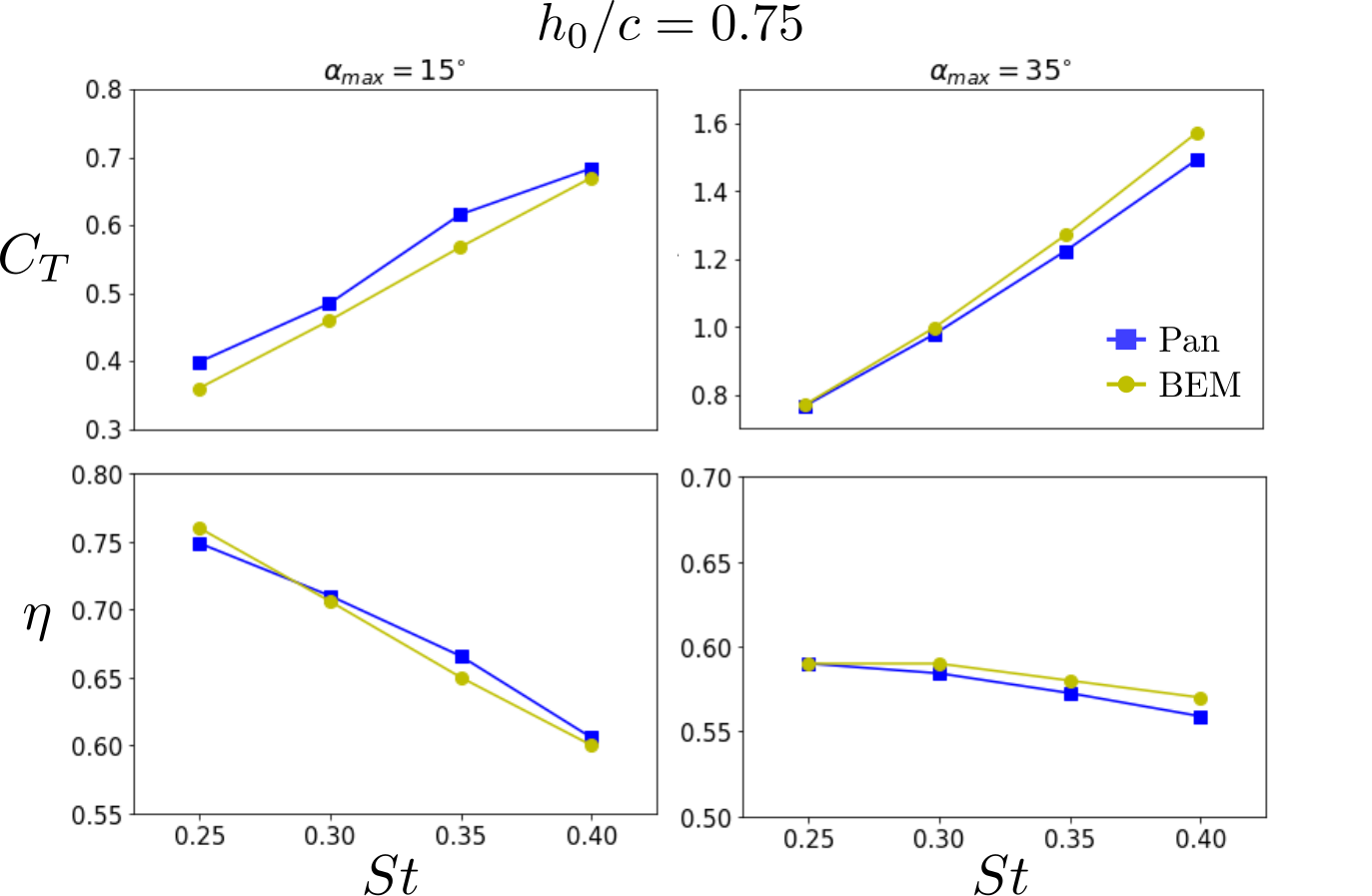}
    \caption{Comparison of the numerical results of the potential flow solver in \S\ref{S:flow_bem} with the numerical solutions by Pan \textit{et al}. \cite{pan2012boundary}. A combined heaving and pitching motion with a heave-to-chord ratio of $h_0/c=0.75$ reaching maximum angles of attack of $15^{\circ}$ (left) and $35^{\circ}$ (right)  are performed over Strouhal numbers ranging from 0.25 to 0.4. The results are compared with respect to the coefficient of thrust $C_T$ (top) and efficiency $\eta$ (bottom). The blue squares represent the work of Pan \textit{et al.} and the yellow circles represent potential flow solver in this work.}
    \label{fig:pan_comparison}
\end{figure}

The theory of Garrick is applicable to low-amplitude motion, and additional validation against large amplitude motions is necessary to give confidence in the potential flow solver. The work of Pan \textit{et al}. \cite{pan2012boundary}  developed a boundary element method to investigate leading-edge separation of heaving and pitching foils, but that is not necessary for the work presented here as we assume that the flow remains attached over the propulsive surface being studied. In the work of Pan \textit{et al}., thrust and efficiency are found for heaving and pitching foils without leading-edge separation for a range of Strouhal numbers and maximum angles of attack $\alpha_{\rm{max}}$. Figure \ref{fig:pan_comparison} shows good agreement between the method presented here and the work of Pan \textit{et al}. for a heave-to-chord ratio of $0.75$ and $\alpha_{\rm{max}} = 35^{\circ}$ over $0.25 < St < 0.4$. 

\section{}\label{s:plane_wave}
The capability of the acoustic boundary element method to model acoustic scattering by a solid body is demonstrated and validated. A rigid circle of radius $a$ placed at the origin that is bombarded by a harmonic field of plane waves. The incident field of unit strength has the form 
\begin{equation*}
P_{\rm{i}}(x,t) =  \exp[i(\kappa r \cos\theta -\omega t)],
\end{equation*}
where $\omega$ is the angular frequency, $\kappa$ is the wavenumber, and $x = r\cos\theta$. The analytical result for the scattered field is \cite{junger1986sound}
\begin{equation}\label{eqn:plane_wave_scattered}
P_{\rm{s}}(x,t) =  e^{i\omega t}\sum_{n=0}^\infty \epsilon_ni^n\left[J_n(\kappa a) -\frac{J'_n(\kappa a)H_n(\kappa r)}{H'_n(\kappa a)}\cos n\theta\right].
\end{equation}
The total acoustic field is the sum of the incident and scattered fields, $P_{\rm{t}} = P_{\rm{s}} + P_{\rm{i}}.$

\begin{figure}
\includegraphics[width =\columnwidth]{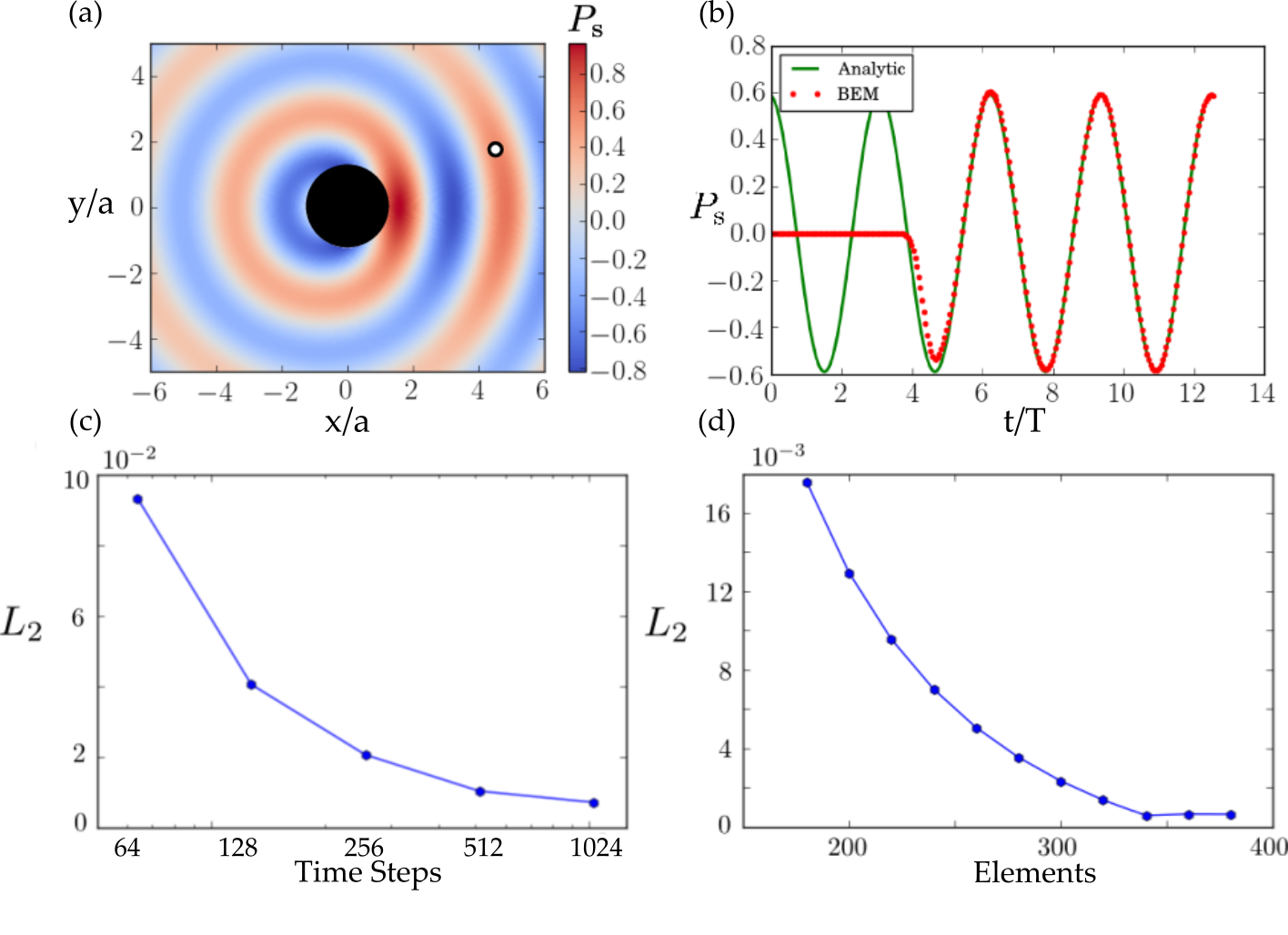}
\caption{A comparison of the analytical to BEM results of the plane wave scatterer study with convergence studies. $(a)$ shows the fully developed scattered field. The observation point, denoted by a black circle,  is placed at the arbitrary point $(r,\theta) = (5,\frac{\pi}{9})$. $(b)$ compares the time history of the scattered field at the observation point for $\omega = 1$ and $\kappa = 2$ with the analytic solution. $(c)$ shows the spatial convergence of the solution, while $(d)$ shows the temporal convergence of the solution}
\label{f:planewaveCollage}
\end{figure}

The interaction of the harmonic incident field with the solid cylinder is as follows. The incoming plane waves propagate in the positive $x$-direction and make initial contact with the cylinder at $(r,\theta) = (a,\pi)$. In the area in front of the cylinder, the plane waves are reflected back onto themselves. The waves reflect at the front of the cylinder to create a shadow region aft of the body. The length of the  shadow region  is dictated by the wavenumber, with larger values resulting in a smaller shadow region.  The $L_2$ error norm is calculated over observation points placed on five circles of five points, sampling all of the regions of the scattered field from distances of $1a \rightarrow 10a$ from the rigid circle. 

Figure~\ref{f:planewaveCollage} $(a)$ compares the transient acoustic response at a point in the acoustic field to the analytical solution to harmonic wave forcing. Here $\omega=1$, $\kappa = 2$, and arbitrary point $(r,\theta) = (5,\frac{\pi}{9})$ are selected for this example. Note the absence of a signal in the BEM solution until the initial scattered wave reaches the observation point, after which the numerical solution quickly converges to the analytical result.

Temporal and spatial discretization independence of the numerical solution are shown in figure \ref{f:planewaveCollage} $(c)$ and $(d)$. For the spatial convergence study, four periods of T $= \pi$ are divided into 256 equidistant time-steps. An increasing number of elements on the boundary were used to compare the BEM solution with (\ref{eqn:plane_wave_scattered}). The temporal convergence study (figure \ref{f:planewaveCollage} $(d)$) had a boundary of 1024 equal length elements over a total period of T $= 4\pi$. The total period is divided into increasing numbers of equidistant time steps. Spatial convergence occurs at approximately 512 elements, showing a relative error of less than 0.1\% when using more than 256 time steps.

\bibliographystyle{bibstyle}
\bibliography{fishModels.bib}

\end{document}